# Odd Viscodiffusive Fluids

**Alhad Deshpande**[a], **Cory Hargus**[a,b], **Karthik Shekhar**[a,c,1], **and Kranthi K. Mandadapu**[a,d,2]

[a]Department of Chemical and Biomolecular Engineering, University of California, Berkeley, California 94720, USA; [b]Université Paris Cité, Laboratoire Matière et Systèmes Complexes (MSC), UMR 7057 CNRS, F-75205, 75205 Paris, France; [c]Helen Wills Neuroscience Institute, University of California, Berkeley, California 94720, USA; [d]Chemical Sciences Division, Lawrence Berkeley National Laboratory, Berkeley, California 94720, USA

We introduce a theory of "odd viscodiffusive fluids," which exhibit three-dimensional odd transport phenomena through the coupling of viscous and diffusive transport. In these fluids, diffusive fluxes may arise from orthogonal velocity gradients and, reciprocally, stresses may arise from concentration gradients. We examine microscopic fluctuations using the recently proposed "flux hypothesis" to derive Green-Kubo and reciprocal relations for the governing transport coefficients. These relations suggest that only parity symmetry, and not time-reversal symmetry, must be broken at the microscopic scale to observe these couplings. Chiral liquids, whether passive or active, are therefore a natural choice as viscodiffusive fluids. We then introduce two analytically tractable model systems, namely a generator and a corresponding reciprocal engine, which illustrate the nature of viscodiffusive cross-coupling in chiral matter and enable the experimental measurement of the novel transport coefficients. Finally, we make the case for chiral bacterial suspensions to be odd viscodiffusive fluids, and use our theory to predict the behaviors exhibited in prior experimental microfluidic studies involving bacterial migration in response to shearing flows.

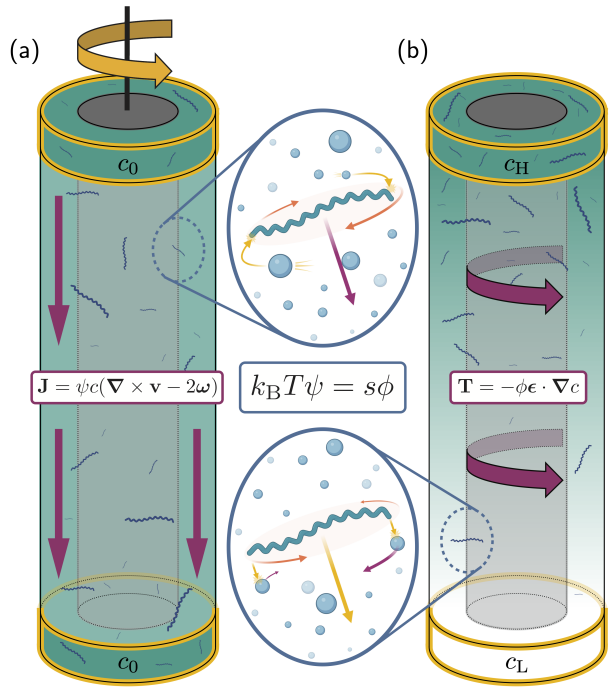

**Fig. 1.** A model odd viscodiffusive fluid consisting of a mixture of chiral solutes in an achiral solvent between two concentric cylinders. **(a)** Chiral generator: An imposed mechanical rotation (yellow) of the inner cylinder, with the outer cylinder fixed, generates a diffusive flux **J** (purple) of the chiral solutes. Inset: a schematic depicting the solute/solvent collisions that cause the solutes to spin and drift orthogonally to the induced shear flow. **(b)** Chiral engine: An applied concentration difference $(c_\mathrm{H} - c_\mathrm{L})$ (yellow) of chiral solutes creates an antisymmetric stress **T**, which causes rotation of the inner cylinder (purple), while the outer cylinder is held fixed. Inset: as solutes migrate (yellow) down the concentration gradient, solute/solvent collisions induce solute spin and vortical motion of surrounding solvent (purple). The reciprocity in the microscopic dynamics of the chiral solutes manifests as reciprocity of the macroscopic transport coefficients $\psi$ and $\phi$, scaled by the thermal energy $k_\mathrm{B}T$ and solute structure factor $s$. The exact nature of how the reciprocal effects emerge in these model systems (a) and (b) will be quantitatively addressed in later sections.

The recognition that microscopic symmetries impose fundamental constraints on macroscopic phenomena dates back at least as far as 1894, when Pierre Curie proposed his symmetry principle (1), providing a generic link between the symmetry of microscopic "causes" and macroscopic "effects." In the realm of transport phenomena, the symmetries exhibited by microscopic fluctuations manifest at the level of constitutive behavior, as formalized by Onsager (2, 3). In recent years, growing interest in systems which break certain microscopic symmetries has motivated a reexamination of known transport phenomena and initiated a search for new ones. Odd, or parity symmetry (PS) breaking, transport phenomena coupling fluxes to spatially-orthogonal driving forces (e.g. gradients) offer a particularly interesting example, having been identified in numerous chiral active systems (4–7). Odd transport involving *conjugate* flux-gradient pairs such as odd viscosity (8–11), odd diffusion (12–14), and odd thermal conduction (15) have been observed in two-dimensional (2D) isotropic systems that break time-reversal symmetry (TRS), in other words microscopic reversibility, in addition to PS at the microscale. By isotropy, however, these phenomena cannot exist in three-dimensional (3D) systems (16), raising the question of whether isotropic odd transport is strictly a 2D phenomenon. However, it has long been known that fluxes may in general arise from driving forces beyond their own conjugates, as in the Dufour and Soret effects (17) which relate heat fluxes to concentration gradients and mass fluxes to temperature gradients. Can *non-conjugate* flux-gradient couplings then give rise to isotropic odd transport in 3D?

In this article, we introduce a theory of "odd viscodiffusive fluids" that exhibit 3D odd transport through the non-conjugate coupling of diffusive fluxes to velocity gradients and, conversely, of stresses to concentration gradients. Such viscodiffusive cross-couplings are traditionally neglected as they supposedly violate the Curie symmetry principle (1), which many have misconstrued that flux-gradient pairs of unalike tensorial rank do not couple (17–22). On the contrary, we show here that viscodiffusive couplings do generically exist in 3D PS-breaking (chiral) fluids. Moreover, unlike other odd transport phenomena (12, 15, 16), odd viscodiffusive transport occurs in both passive (TRS-preserving) and active (TRS-breaking) systems. For the former, we develop a reciprocal relation (2, 3) that unifies viscodiffusive transport with a single transport coefficient. We illustrate the macroscopic nature of these fluids by considering two reciprocal model systems, namely a "chiral generator" and a "chiral engine" as depicted in Fig. 1. Notably, the former induces solute diffusive fluxes by imposing mechanical rotation, while the latter generates mechanical motion when maintaining a concentration gradient. As odd viscodiffusive fluids are by nature PS-breaking, solutions of chiral objects become natural candidates for their experimental realization. In particular, we analyze recent microfluidic experiments (23) involving solutions of helical bacteria (spirochetes) that drift orthogonally to a shear plane under Poiseuille flow—and in doing so measure the single governing viscodiffusive transport coefficient. We end with a discussion of the Curie symmetry principle (1), where we clarify the original statement and demonstrate its consistency with our work.

## Transport of Odd Viscodiffusive Fluids

**Balance Laws.** We begin by writing balance laws and constitutive relations for a nonreacting binary fluid solution composed of a solute and a solvent. We take both components to be rigid particles. The system may break parity symmetry (PS) through the geometric chirality of either component, or through the chirality of interparticle interactions. We may write balance laws for the solute concentration $c$, linear momentum density $\rho\mathbf{v}$, and intrinsic angular momentum density (spin field) $\rho\mathbf{M}$, with $\rho$ being the fluid mass density, as

$$\dot{c} + c\boldsymbol{\nabla}\cdot\mathbf{v} = -\boldsymbol{\nabla}\cdot\mathbf{J}, \quad [1]$$

$$\rho\dot{\mathbf{v}} = \rho\mathbf{b} + \boldsymbol{\nabla}\cdot\mathbf{T}, \quad [2]$$

$$\rho\dot{\mathbf{M}} = \rho\mathbf{G} + \boldsymbol{\nabla}\cdot\mathbf{C} - \boldsymbol{\epsilon}:\mathbf{T}, \quad [3]$$

where $(\dot{\ }) := \frac{\partial}{\partial t} + \mathbf{v}\cdot\boldsymbol{\nabla}$ is the material derivative (17, 24). The diffusive flux $\mathbf{J}$, stress tensor $\mathbf{T}$, and couple stress tensor $\mathbf{C}$ refer to fluxes of solute, linear momentum, and spin angular momentum, respectively; $\rho\mathbf{b}$ and $\rho\mathbf{G}$, respectively, denote body forces and body torques. Notably, we consider the spin evolution Eq. [3] in anticipation that odd viscodiffusive effects involve asymmetry of the stress tensor $\mathbf{T}$. The antisymmetric components of the stress are defined as $\boldsymbol{\epsilon}:\mathbf{T} := \epsilon_{ijk}T_{jk}\mathbf{e}_i$, where $\epsilon_{ijk}$ is the Levi-Civita tensor and Einstein summation notation is assumed. In systems where there is no interconversion between spin and orbital angular momentum, Eq. [3] reduces to the symmetry of the stress tensor $\boldsymbol{\epsilon}:\mathbf{T} = \mathbf{0}$ (24).

**Constitutive Relations.** We now present linear constitutive laws relating the fluxes $\mathbf{J}, \mathbf{T}$, and $\mathbf{C}$ to driving forces. In addition to each flux's conjugate driving force (17), we allow for non-conjugate couplings; namely, we consider a viscodiffusive flux $\mathbf{J}$ arising from a velocity gradient $\boldsymbol{\nabla}\mathbf{v}$ and spin field $\mathbf{M}$, and a viscodiffusive stress $\mathbf{T}$ arising from a concentration gradient $\boldsymbol{\nabla}c$. We consider the case where the couple stress $\mathbf{C}$ depends only on its conjugate[*], i.e., the spin gradient $\boldsymbol{\nabla}\mathbf{M}$. The most general linear relations are then the tensorial linear maps between the fluxes and the driving forces. Invoking isotropy for 3D systems, each map can be expressed as a linear combination of the rank two Kronecker delta tensor $\boldsymbol{\delta}$ and the rank three Levi-Civita tensor $\boldsymbol{\epsilon}$ (16, 25, 26). Thus, the mappings are simplified and we obtain the following constitutive relations (see Appendix A for more details):

$$\mathbf{J} = -D\boldsymbol{\nabla}c + \psi c\boldsymbol{\nabla}\times\mathbf{v} - \eta c\mathbf{M}, \quad [4]$$

$$\mathbf{T} = -p\boldsymbol{\delta} + (\lambda_1 - \frac{2}{3}\lambda_2)(\boldsymbol{\nabla}\cdot\mathbf{v})\boldsymbol{\delta} + \lambda_2(\boldsymbol{\nabla}\mathbf{v} + \boldsymbol{\nabla}\mathbf{v}^\mathsf{T}) \\ + \lambda_3(\boldsymbol{\nabla}\mathbf{v} - \boldsymbol{\nabla}\mathbf{v}^\mathsf{T}) + \gamma\boldsymbol{\epsilon}\cdot\mathbf{M} - \phi\boldsymbol{\epsilon}\cdot\boldsymbol{\nabla}c, \quad [5]$$

$$\mathbf{C} = -p_C\boldsymbol{\delta} + (\alpha_1 - \frac{2}{3}\alpha_2)(\boldsymbol{\nabla}\cdot\mathbf{M})\boldsymbol{\delta} \\ + \alpha_2(\boldsymbol{\nabla}\mathbf{M} + \boldsymbol{\nabla}\mathbf{M}^\mathsf{T}) + \alpha_3(\boldsymbol{\nabla}\mathbf{M} - \boldsymbol{\nabla}\mathbf{M}^\mathsf{T}). \quad [6]$$

Viscodiffusive effects stem from the pseudoscalar transport coefficients $\psi$, $\eta$, and $\phi$, which are odd under parity symmetry and with signs corresponding to the chirality of the system.[‡] These coefficients are notably absent in 2D isotropic systems as no rank three isotropic tensor exists (16)—therefore, odd viscodiffusive effects are exclusively 3D phenomena for isotropic systems. The coefficients $\psi$ and $\eta$ relate vorticity and spin, respectively, to the diffusive flux as shown in Fig. 1(a). Through the reciprocal effect, $\phi$ relates concentration gradients to antisymmetric stresses, as shown in Fig. 1(b). The remaining transport coefficients in Eqs. [4]-[6] are commonplace in classical transport phenomena (17, 27, 28): $D$ is the diffusion coefficient, $\lambda_1$ is the dilatational viscosity, $\lambda_2$ is the shear viscosity, $\lambda_3$ is the rotational viscosity, $\alpha_1$ is the dilatational spin viscosity, $\alpha_2$ is the shear spin viscosity, and $\alpha_3$ is the rotational spin viscosity. The coefficient $\gamma$, which relates the spin field to the stress, is proportional to the rotational viscosity $\lambda_3$, as will be shown in a later section. Moreover, the stress tensor $\mathbf{T}$ and couple stress tensor $\mathbf{C}$ may possess hydrostatic terms, respectively termed the pressure $p$ and couple pressure $p_C$.

For constant transport coefficients, the constitutive relations [4]-[6] in conjunction with the balance laws [1]-[3] yield the evolution equations,

$$\dot{c} + c\boldsymbol{\nabla}\cdot\mathbf{v} = D\nabla^2 c - (\psi\boldsymbol{\nabla}\times\mathbf{v} - \eta\mathbf{M})\cdot\boldsymbol{\nabla}c + c\boldsymbol{\nabla}\cdot\eta\mathbf{M}, \quad [7]$$

$$\rho\dot{\mathbf{v}} = \rho\mathbf{b} - \boldsymbol{\nabla}p + (\lambda_1 + \frac{1}{3}\lambda_2)\boldsymbol{\nabla}(\boldsymbol{\nabla}\cdot\mathbf{v}) \\ + \lambda_2\nabla^2\mathbf{v} - \boldsymbol{\nabla}\times(\lambda_3\boldsymbol{\nabla}\times\mathbf{v} - \gamma\mathbf{M}), \quad [8]$$

$$\rho\dot{\mathbf{M}} = \rho\mathbf{G} - \boldsymbol{\nabla}p_C + (\alpha_1 + \frac{1}{3}\alpha_2 - \alpha_3)\boldsymbol{\nabla}(\boldsymbol{\nabla}\cdot\mathbf{M}) \\ + (\alpha_2 + \alpha_3)\nabla^2\mathbf{M} + 2(\lambda_3\boldsymbol{\nabla}\times\mathbf{v} - \gamma\mathbf{M} + \phi\boldsymbol{\nabla}c), \quad [9]$$

where $\nabla^2$ is the Laplacian operator. As can be seen from Eq. [7], the viscodiffusive coefficients $\psi$ and $\eta$ contribute to advective transport of the solute with propagation velocity $(\psi\boldsymbol{\nabla}\times\mathbf{v} - \eta\mathbf{M})$. In addition, the concentration field may evolve through a source-like term arising from the divergence of the spin field. Notably, the odd transport coefficient $\phi$ does not enter the "extended" Navier-Stokes equation [8] (17, 29), but may influence the velocity field through boundary conditions or through the spin balance [9] by way of the asymmetry of the stress tensor $\mathbf{T}$. We shall argue in later sections that the overall stress asymmetry is crucial for the manifestation of viscodiffusive effects on a macroscopic scale.

---

[*] The scope of this work involves a simplification of the more general case where the fluxes $\mathbf{J}$ and $\mathbf{T}$ may depend on the higher order contribution of the spin gradient $\boldsymbol{\nabla}\mathbf{M}$ and reciprocally, the couple stress $\mathbf{C}$ may depend on the concentration gradient $\boldsymbol{\nabla}c$, velocity gradient $\boldsymbol{\nabla}\mathbf{v}$, and spin field $\mathbf{M}$.

[‡] Without loss of generality, a factor of $c$ accompanies the coefficients $\psi$ and $\eta$ as the diffusive flux $\mathbf{J}$ must be linearly proportional to the concentration $c$ to the lowest order.

## Microscopic Fluctuations & Parity Symmetry Breaking

**Flux Hypothesis & Green-Kubo Relations.** Odd transport phenomena such as odd diffusion (12) and odd viscosity (9, 16) have been shown to require the breaking of TRS and PS at the microscale. We ask what then are the microscopic requirements to observe odd viscodiffusive transport, or equivalently, when do $\psi, \eta, \phi \neq 0$? Such requirements are best understood by examining microscopic fluctuations through Green-Kubo relations.

Green-Kubo relations connect macroscopic transport coefficients to time correlation functions of microscopic fluctuations, thereby demonstrating the explicit roles PS and TRS play in observing certain phenomena. In addition to providing a statistical mechanical foundation for continuum transport phenomena, they enable for the measurement of transport coefficients from molecular simulations. Classically, these relations are derived from the Onsager regression hypothesis (2, 3) which states that the average regression of small microscopic fluctuations from a steady state follow the same laws as the temporal relaxation of macroscopic field variables when perturbed externally. As utilized by Kubo, Yokota, and Nakajima (30), the hypothesis acts at the level of the balance laws [7]-[9]. However, the viscodiffusive coefficients $\psi, \eta$, and $\phi$ do not enter the balance laws through *linear* (small) perturbations about steady states, and thus are incapable of being analyzed through the regression hypothesis (31). Recent work addresses this issue by proposing a "flux hypothesis" (31), a generalization of the regression hypothesis. The flux hypothesis acts at the level of the constitutive relations [4]-[6] and states that microscopic fluxes arising from spontaneous fluctuations on average follow macroscopic constitutive behavior. In this work, starting from the flux hypothesis we develop Green-Kubo relations for each transport coefficient presented in Eqs. [4]-[6], which hold for both active and passive systems. In what follows, we present only the Green-Kubo relations for the viscodiffusive coefficients $\psi, \eta$, and $\phi$, while all other relations and associated derivations are available in Appendix B.

We consider microscopic fluctuations about a spatially homogeneous, possibly non-equilibrium, steady-state. For instance, $\delta\hat{A}(\mathbf{x}, t) = \hat{A}(\mathbf{x}, t) - \langle \hat{A} \rangle$ refers to an instantaneous microscopic fluctuation corresponding to the macroscopic field $A$, where $\langle \cdot \rangle$ denotes a steady-state ensemble average. Defining the Fourier transform as $\delta\hat{A}^{\mathbf{q}}(t) := \int d\mathbf{x} e^{-i\mathbf{q}\cdot\mathbf{x}} \delta\hat{A}(\mathbf{x}, t)$ with wave vector $\mathbf{q}$, we obtain the Green-Kubo relations in terms of time correlations between the stress and diffusive flux as

$$\psi = -\frac{\epsilon_{ijk}(1-\tau/\nu)}{6\rho_0 c_0 \mu V} \lim_{\mathbf{q}\to\mathbf{0}} \int_0^{\Delta t} dt \left\langle \delta\hat{J}_i^{\mathbf{q}}(t) \delta\hat{T}_{jk}^{-\mathbf{q}}(0) \right\rangle, \quad [10]$$

$$\eta = -\frac{\epsilon_{ijk}}{3\rho_0 c_0 \nu V} \lim_{\mathbf{q}\to\mathbf{0}} \int_0^{\Delta t} dt \left\langle \delta\hat{J}_i^{\mathbf{q}}(t) \delta\hat{T}_{jk}^{-\mathbf{q}}(0) \right\rangle, \quad [11]$$

$$\phi = \frac{\epsilon_{ijk}}{6sc_0 V} \lim_{\mathbf{q}\to\mathbf{0}} \int_0^{\Delta t} dt \left\langle \delta\hat{T}_{ij}^{\mathbf{q}}(t) \delta\hat{J}_k^{-\mathbf{q}}(0) \right\rangle. \quad [12]$$

Here $\rho_0$ denotes the bulk mass density, $c_0$ denotes the bulk solute concentration, and the static correlators $s, \mu, \nu$, and $\tau$ are defined as

$$s := \frac{1}{c_0 V} \lim_{\mathbf{q}\to\mathbf{0}} \left\langle \delta\hat{c}^{\mathbf{q}} \delta\hat{c}^{-\mathbf{q}} \right\rangle, \quad [13]$$

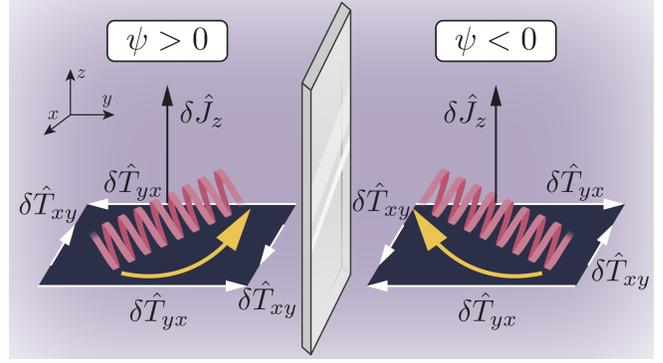

**Fig. 2.** A schematic of 3D parity symmetry (PS) breaking fluctuations correlating the diffusive flux and antisymmetric stress encoded in the Green-Kubo relation Eq. [10]. For a fluid consisting of helical solutes, fluctuations in the flux $\delta\hat{J}_z$ are correlated with fluctuations in the antisymmetric stress $\delta\hat{T}_{xy} - \delta\hat{T}_{yx}$, the latter creating a net torque. Here, assuming that $|\delta\hat{T}_{yx}| > |\delta\hat{T}_{xy}|$, this net torque is indicated by the yellow arrow, and on average provides a propensity for a right-handed helix to move upward (20, 23), yielding a positive $\delta\hat{J}_z$ and therefore obtaining $\psi > 0$. Upon the mirror inversion $x \mapsto x$, $y \mapsto -y$, $z \mapsto z$, $\delta\hat{T}_{xy} - \delta\hat{T}_{yx}$ changes sign, while $\delta\hat{J}_z$ remains invariant since the helix changes its handedness. Accordingly, $\psi$ changes sign, thereby making it odd under PS. The schematic also applies to understand the PS-breaking correlations in the Green-Kubo relations [11] and [12] for the other viscodiffusive coefficients $\eta$ and $\phi$.

$$\mu\delta_{ij} := \frac{1}{V} \lim_{\mathbf{q}\to\mathbf{0}} \left\langle \delta\hat{v}_i^{\mathbf{q}} \delta\hat{v}_j^{-\mathbf{q}} \right\rangle, \quad [14]$$

$$\nu\delta_{ij} := \frac{1}{V} \lim_{\mathbf{q}\to\mathbf{0}} \left\langle \delta\hat{M}_i^{\mathbf{q}} \delta\hat{M}_j^{-\mathbf{q}} \right\rangle, \quad [15]$$

$$\tau = \frac{1}{3V} \lim_{\mathbf{q}\to\mathbf{0}} \epsilon_{ijk} \frac{\partial}{\partial iq_i} \left\langle \delta\hat{M}_j^{\mathbf{q}} \delta\hat{v}_k^{-\mathbf{q}} \right\rangle. \quad [16]$$

The Green-Kubo relations [10]-[12] recover the macroscopic transport coefficients in the $\mathbf{q} \to \mathbf{0}$ large wavelength limit. Furthermore, they are valid only under a separation of timescales such that $\Delta t$ is sufficiently small compared to the macroscopic relaxation time, but large compared to the decay of the time correlation functions. The static correlator $s$ is the solute structure factor, and the static correlators $\mu$ and $\nu$ are measures of translational and rotational kinetic temperatures. In the special case of equilibrium, i.e. passive, systems, velocity and spin fluctuations are statically decorrelated and proportional through equipartitioning, where $\tau = 0$ and $\rho_0\mu = \rho_0^2\nu/I_0 = k_\mathrm{B}T$, with moment of inertia density $I_0$, Boltzmann constant $k_\mathrm{B}$ and temperature $T$.

Further, Eqs. [10]-[12] demonstrate that viscodiffusive fluxes arise from the cross correlations of antisymmetric stress fluctuations and solute flux fluctuations. The presence of the axial tensor $\epsilon_{ijk}$ in these relations points to the PS-breaking nature of the transport coefficients, as depicted in Fig. 2 for a fluid composed of (chiral) helices. Taking the mirror inversion of the system alters the chirality of the fluid, which changes the sign of the coefficient $\psi$ (and in turn of $\eta$ and $\phi$). In addition, the Green-Kubo relations [10]-[12] indicate that the viscodiffusive coefficients contain both TRS-preserving and TRS-breaking contributions, allowing them to exist in both passive and active systems [§]. Thus, unlike odd diffusion (12) and odd viscosity (9, 16), TRS-breaking is not a microscopic requirement to observe odd viscodiffusive transport. The minimum determining feature of viscodiffusive fluids is then the breaking of PS, i.e. the presence of chirality. A summary of the roles of PS and TRS for each transport coefficient, along with the role of asymmetric stress fluctuations is provided in Table 1. Lastly, the Green-Kubo relations [10], [11], [97], and [105] also reveal the coefficients coupling vorticity and spin to diffusive flux and to stress obey the relation

$$\frac{2\psi}{\eta} = \frac{2\lambda_3}{\gamma} = \frac{\nu - \tau}{\mu}, \qquad [17]$$

whether TRS is preserved or not.

---

[§] In contrast, all other transport coefficients in Eqs. [4]-[6] contain only TRS-preserving contributions. Therefore, even in active systems, the values of the non-viscodiffusive coefficients will be invariant to time-reversal.

**Table 1. Symmetries of transport coefficients.** The first column indicates the transport coefficient, while the second and third columns indicate the associated flux and driving force. The fourth and fifth columns indicate the even (+) and odd (−) nature of the transport coefficients upon parity and time reversal transformations. The viscodiffusive coefficients $\psi$, $\eta$ and $\phi$ contain both TRS-preserving and TRS-breaking contributions and thus are marked $+/-$ under time reversal. The last column indicates whether coefficients may be nonzero (✓) or are 0 in systems with strictly symmetric stress *fluctuations*, showing that stress asymmetry is necessary for the existence of nonzero odd viscodiffusive coefficients. Note that in a more general case, for systems with couple stress–antisymmetric stress correlations, the spin viscosities $\alpha_1, \alpha_2,$ and $\alpha_3$ are $+/-$ under parity and time reversal transformations.

| Coefficient | Flux | Force | PS | TRS | $\delta\hat{\mathbf{T}} = \delta\hat{\mathbf{T}}^\mathsf{T}$ |
|---|---|---|---|---|---|
| $D$ | | $\boldsymbol{\nabla} c$ | + | + | ✓ |
| $\psi$ | $\mathbf{J}$ | $\boldsymbol{\nabla}\mathbf{v}$ | − | +/− | 0 |
| $\eta$ | | $\mathbf{M}$ | − | +/− | 0 |
| $\phi$ | | $\boldsymbol{\nabla} c$ | − | +/− | 0 |
| $\lambda_1$ | | $\boldsymbol{\nabla}\mathbf{v}$ | + | + | ✓ |
| $\lambda_2$ | $\mathbf{T}$ | $\boldsymbol{\nabla}\mathbf{v}$ | + | + | ✓ |
| $\lambda_3$ | | $\boldsymbol{\nabla}\mathbf{v}$ | + | + | 0 |
| $\gamma$ | | $\mathbf{M}$ | + | + | 0 |
| $\alpha_1$ | | | + | + | ✓ |
| $\alpha_2$ | $\mathbf{C}$ | $\boldsymbol{\nabla}\mathbf{M}$ | + | + | ✓ |
| $\alpha_3$ | | | + | + | ✓ |

**Passive Viscodiffusive Fluids and Reciprocal Relation.** As odd viscodiffusive fluids do not require TRS-breaking, they may be either passive or active. Considering the case of passive fluids that are in thermal equilibrium when unperturbed, TRS or microscopic reversibility (2) dictates that the cross-correlations between the diffusive flux and stress in the Green-Kubo relations satisfy

$$\left\langle \delta\hat{J}_i^\mathbf{q}(t)\delta\hat{T}_{jk}^{-\mathbf{q}}(0) \right\rangle \stackrel{\text{TRS}}{=} -\left\langle \delta\hat{J}_i^\mathbf{q}(0)\delta\hat{T}_{jk}^{-\mathbf{q}}(t) \right\rangle. \qquad [18]$$

Here, the minus sign is attained as the stress and diffusive flux are, respectively, even and odd functions of the particle velocities (19, 32). Furthermore, in equilibrium, $\rho_0\mu = k_\mathrm{B}T$ and $\tau = 0$ by the equipartition theorem (17). Given the condition of microscopic reversibility [18], the Green-Kubo relations for the viscodiffusive cross-coupling coefficients $\psi$ and $\phi$ in [10] and [12] then yield the reciprocal relation

$$\phi \stackrel{\text{equil}}{=} \frac{k_\mathrm{B}T}{s}\psi \stackrel{\text{dilute}}{\approx} k_\mathrm{B}T\psi, \qquad [19]$$

where the solute structure factor $s = 1$ in sufficiently dilute systems. For isotropic equilibrium systems, the reciprocal response to a diffusive flux arising from a vorticity is then an antisymmetric stress arising from a concentration gradient, as shown schematically in the insets in Fig. 1.

For isotropic systems, we may express the spin angular momentum density as $\rho\mathbf{M} = I\boldsymbol{\omega}$, with $I$ being the scalar moment of inertia and $\boldsymbol{\omega}$ the angular velocity field. In equilibrium, $\rho_0\nu = k_\mathrm{B}T(I_0/\rho_0)$ by equipartition, which therefore reduces Eq. [17] to

$$\frac{2\psi}{\eta} = \frac{2\lambda_3}{\gamma} \stackrel{\text{equil}}{=} \frac{\nu}{\mu} \stackrel{\text{equil}}{=} \frac{I_0}{\rho_0}. \qquad [20]$$

Together, the relations in [20] and the reciprocal relation [19] reduce the three viscodiffusive coefficients $\psi$, $\eta$, and $\phi$ to a single coefficient. These relations modify the constitutive laws [4] and [5] to

$$\mathbf{J} \stackrel{\text{equil}}{=} -D\boldsymbol{\nabla} c + \psi c(\boldsymbol{\nabla}\times\mathbf{v} - 2\boldsymbol{\omega}), \qquad [21]$$

$$\mathbf{T} \stackrel{\text{equil}}{=} -p\boldsymbol{\delta} + (\lambda_1 - \frac{2}{3}\lambda_2)(\boldsymbol{\nabla}\cdot\mathbf{v})\boldsymbol{\delta} + \lambda_2(\boldsymbol{\nabla}\mathbf{v} + \boldsymbol{\nabla}\mathbf{v}^\mathsf{T})$$
$$- \lambda_3\boldsymbol{\epsilon}\cdot(\boldsymbol{\nabla}\times\mathbf{v} - 2\boldsymbol{\omega}) - \frac{k_\mathrm{B}T}{s}\psi\boldsymbol{\epsilon}\cdot\boldsymbol{\nabla} c. \qquad [22]$$

Hence, it is solely the difference in vorticity and angular velocity, $\boldsymbol{\nabla}\times\mathbf{v} - 2\boldsymbol{\omega}$, sometimes referred to as the "sprain rate" (33), that drives fluxes in equilibrium systems, preserving invariance to rigid body rotation. While constitutive laws are classically proposed directly relating the sprain rate to a stress (17, 29), in active systems, fluxes may individually depend on vorticity and spin (16).

**Implication to Macroscopic Stress Asymmetry.** The Green-Kubo relations [10]-[12] demonstrate the requirement of *microscopic* asymmetric stress fluctuations in observing nonzero viscodiffusive transport coefficients. We now ascertain the role of stress asymmetry in viscodiffusive transport at a *macroscale* for both passive and active fluids. If, for instance, the stress is symmetric ($\boldsymbol{\epsilon}:\mathbf{T} = \mathbf{0}$), relation [17] reduces the constitutive laws [4] and [5] to

$$\mathbf{J} \stackrel{\text{sym}}{=} -\left(D + \frac{\psi\phi}{\lambda_3}c\right)\boldsymbol{\nabla} c \stackrel{\text{dilute}}{\approx} -D\boldsymbol{\nabla} c, \qquad [23]$$

$$\mathbf{T} \stackrel{\text{sym}}{=} -p\boldsymbol{\delta} + (\lambda_1 - \frac{2}{3}\lambda_2)(\boldsymbol{\nabla}\cdot\mathbf{v})\boldsymbol{\delta} + \lambda_2(\boldsymbol{\nabla}\mathbf{v} + \boldsymbol{\nabla}\mathbf{v}^\mathsf{T}). \qquad [24]$$

corresponding to standard forms of Fick's and Newton's laws observed in achiral 3D systems (17, 27, 28). In this case, viscodiffusive effects only manifest as a nonlinear addition to the diffusion coefficient scaled by $\psi\phi c/\lambda_3$, negligible for dilute systems. Thus, if the stress tensor is symmetric, there will be no net viscodiffusive effects (at least for dilute systems). While microscopic asymmetric stress fluctuations are needed to observe nonzero viscodiffusive coefficients, macroscopic asymmetric stresses are needed to observe the resulting viscodiffusive fluxes at the macroscale.

The intrinsic spin balance [3] provides conditions to observe a symmetric stress tensor. In steady-states with no convective spin transport, antisymmetric stresses exist only in the presence of a body torque $\rho\mathbf{G}$ or a boundary torque corresponding to the term $\boldsymbol{\nabla}\cdot\mathbf{C}$. In the absence of these applied torques, stresses may only be transiently asymmetric as the intrinsic spin field $\mathbf{M}$ adapts to symmetrize the stress at steady-state. In what follows, we analyze a few scenarios illustrating the emergence of viscodiffusive effects, defined by the general linear constitutive laws [4]-[6].

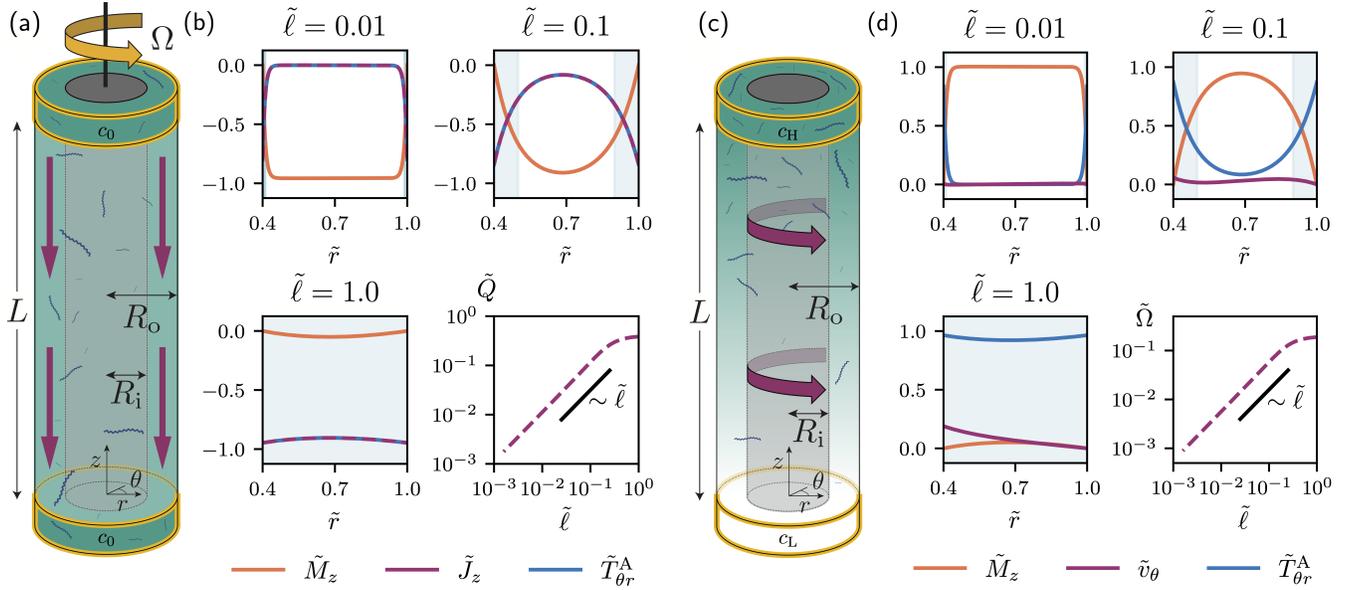

**Fig. 3.** Analysis of the chiral generator and chiral engine. **(a)** Schematic of a chiral generator with chemical baths of equal concentrations to eliminate concentration gradients. **(b)** The dimensionless spin $\tilde{M}_z$, diffusive flux $\tilde{J}_z$, and antisymmetric stress $\tilde{T}^A_{\theta r}$ for three different values of the boundary layer length scale relative to the system size $\tilde{\ell}$. Indicated in blue are the corresponding boundary layer regions, wherein viscodiffusive effects exist. Also shown is the total integrated viscodiffusive current $\tilde{Q}$, which scales linearly with $\tilde{\ell}$. Note that, in this case, $\tilde{J}_z = \tilde{T}^A_{\theta r}$ and thus the associated curves are superimposed. **(c)** Schematic of a chiral engine with chemical baths of unequal concentration driving a downward diffusive flux. **(d)** The dimensionless spin $\tilde{M}_z$, velocity $\tilde{v}_\theta$, and antisymmetric stress $\tilde{T}^A_{\theta r}$ for three different values of the dimensionless boundary layer $\tilde{\ell}$ with the boundary layer regions shaded in blue. The resulting inner cylinder rotation $\tilde{\Omega} = \tilde{v}_\theta(\xi)$ is also shown and scales linearly with $\tilde{\ell}$. In obtaining the profiles, we use fixed values of $\tilde{\lambda}_3 = 0.167$ (34), relevant for water, and $\xi = R_i/R_o = 0.4$.

## Chiral Generator and Engine

We begin by considering a pair of reciprocal model systems involving coaxial cylinders from which viscodiffusive effects may be measured, termed a "chiral generator" and a "chiral engine," as depicted in Fig. 1. In the generator, an external torque rotates the inner cylinder, shearing the fluid and driving viscodiffusive mass fluxes (akin to the rheological separator involving twisted ribbon-like particles proposed in Ref. (20)). In contrast, for the engine, the system is in contact with unequal concentration reservoirs, causing a mass flux and driving viscodiffusive stresses, which produce a torque on the shaft. The generator and engine represent opposite modes of interconverting mechanical and chemical energy.

To analyze the model systems, boundary conditions are essential for solving the governing differential equations [7], [8], and [9]. While boundary conditions for concentration $c$ and velocity $\mathbf{v}$ are well-understood (27, 28), the boundary conditions involving the spin field $\mathbf{M}$ are less apparent. Past simulation work has highlighted the applicability of zero-spin (Dirichlet) boundary conditions $\mathbf{M} = 0$ at a solid-fluid interface (35, 36). Physically, these would stem from particle-wall frictional interactions and depend on the roughness of the wall relative to the particle size. In the absence of these frictional interactions, the wall will no longer enact a torque on a solute. Thus for a very smooth (e.g. greased) surface or for a liquid-vapor interface, the Neumann condition $\mathbf{Cn} = 0$ is more appropriate. In what follows, we employ the zero-spin boundary condition at solid-liquid interfaces.

**Chiral Generator.** We consider a limiting case of the chiral generator, as depicted in Fig. 3(a), involving an incompressible viscodiffusive fluid existing between two concentric cylinders with radii $R_i$ and $R_o$, and length $L$ such that $R_o \ll L$. The inner cylinder is rotated with angular velocity $\Omega$, so that the fluid in contact with the inner cylinder moves at velocity $v_\theta = R_i \Omega$. The outer cylinder is held stationary. Chemical baths with concentration $c_0$ are stationed at $z = 0$ and $z = L$ to maintain a homogeneous concentration throughout the domain. We proceed to solve for the spatial profiles of the bulk fluid velocity $\mathbf{v}$ and the intrinsic spin field $\mathbf{M}$. Although we may solve the two-dimensional problem with both $r$ and $z$ dependence, the assumption $R_o \ll L$ renders the problem effectively one-dimensional.

We assume the fields follow the ansatzes, $\mathbf{v} = v_\theta(r)\mathbf{e}_\theta$ and $\mathbf{M} = M_z(r)\mathbf{e}_z$. The fields are further non-dimensionalized as $\tilde{v}_\theta = v_\theta/(R_i\Omega)$, $\tilde{M}_z = M_z/(\lambda_3 \xi \Omega/\gamma)$, $\tilde{J}_z = J_z/(\psi c_0 \xi \Omega)$, and $\tilde{T}^A_{\theta r} = (T_{\theta r} - T_{r\theta})/(\lambda_3 \xi \Omega)$ with nondimensional radial coordinate $\tilde{r} = r/R_o$ and ratio of radii $\xi := R_i/R_o$. These yield two dimensionless groups: $\tilde{\lambda}_3 = \lambda_3/(\lambda_2 + \lambda_3)$, setting the ratio of the rotational viscosity to the shear viscosity and $\mathrm{Sp} = (\alpha_2 + \alpha_3)/(2\gamma R_o^2)$, measuring the diffusion of angular momentum relative to the consumption/production of moment of linear momentum through the antisymmetric stress. There exists an analytical solution to the problem setup, provided in Appendix C.

The two dimensionless groups yield boundary layers at the inner and outer cylinders where the stress tensor is asymmetric with a nondimensional length scale $\tilde{\ell} = \sqrt{\mathrm{Sp}/(1-\tilde{\lambda}_3)}$. Since there exist no concentration gradients, the nondimensional diffusive flux is exactly equal to the nondimensional antisymmetric stress, $\tilde{J}_z = \tilde{T}^A_{\theta r}$. Therefore, it is within this length scale $\tilde{\ell}R_o$ that viscodiffusive fluxes persist, as macroscopic stress asymmetry is required to observe viscodiffusive effects. Figure 3(b) shows the spin $\tilde{M}_z$, diffusive flux $\tilde{J}_z$ and anti-symmetric stress $\tilde{T}_{\theta r}^{\mathrm{A}}$ profiles for systems of varying boundary layer length scales relative to the system size. As $\tilde{\ell}$ grows, spin diffusion increasingly dominates, thereby resulting in a larger imbalance of spin and vorticity throughout the system. This increases the spatial extent over which the stress is asymmetric, leading to viscodiffusive fluxes throughout the bulk.

The chiral generator also permits the measurement of the coefficient $\psi$. To that end, we estimate the net viscodiffusive current running down the channel $\tilde{Q} = \int_\xi^1 \tilde{J}\tilde{r}d\tilde{r}$, and find it to be linearly proportional to the relative extent of this boundary layer ($\tilde{\ell}$), see Fig. 3(b). In the limit $\tilde{\ell} \to 0$, the analytical solutions yield $\tilde{Q} = -2\xi(1-\tilde{\lambda}_3)\tilde{\ell}/(1-\xi)$, with the total dimensional current being $Q = 2\pi R_o^2 \psi c_0 \xi \Omega \tilde{Q}$. Thus, with knowledge of quantities $\tilde{\lambda}_3$ and $\tilde{\ell}$, measuring the current $Q$ under an imposed rotation $\Omega$ yields an estimate of the novel transport coefficient $\psi$.

**Chiral Engine.** We now consider the reciprocal setup, i.e., the chiral engine as depicted in Fig. 3(c). The geometry is the same as in the case of the chiral generator. However, we now maintain chemical baths of concentrations $c_L$ and $c_H$ at $z = 0$ and $z = L$, respectively, creating a thermodynamic force that causes the solutes to diffuse down the concentration gradient. Furthermore, while the outer cylinder is held fixed, we allow the inner cylinder to rotate freely, with the corresponding force free condition $\mathbf{Tn} = \mathbf{0}$ with unit normal $\mathbf{n} = -\mathbf{e}_r$. The spin boundary conditions at both walls remain the same as in the generator case, i.e., $\mathbf{M} = \mathbf{0}$. Considering $R_o \ll L$ again renders the problem one-dimensional, and therefore we use the following ansatzes: $c = c(z)$, $\mathbf{v} = v_\theta(r)\mathbf{e}_\theta$ and $\mathbf{M} = M_z(r)\mathbf{e}_z$. The dimensionless variables are $\tilde{c} = (c - c_L)/\Delta c$, $\tilde{v}_\theta = v_\theta/[\phi\Delta c R_o/[(\lambda_2 + \lambda_3)L]]$, $\tilde{M}_z = M_z/[\phi\Delta c/(\gamma L)]$, $\tilde{J}_z = J_z/(D\Delta c/L)$, and $\tilde{T}_{\theta r}^{\mathrm{A}} = (T_{\theta r} - T_{r\theta})/(\phi\Delta c/L)$ with nondimensional radial coordinate $\tilde{r} = r/R_o$ and $\Delta c := c_H - c_L$. The details of the analytical solutions in the limit $c_L \gg \Delta c$ are provided in Appendix C.

Identical to the chiral generator, we attain the same dimensionless groups $\tilde{\lambda}_3$ and Sp, and an associated boundary layer $\tilde{\ell}$ over which the stress is asymmetric. Outside the boundary layer, the spin adapts to symmetrize the stress, as indicated in the profiles for spin $\tilde{M}_z$, velocity $\tilde{v}_\theta$, and antisymmetric stress $\tilde{T}_{\theta r}^{\mathrm{A}}$ in Fig. 3(d). Further, for the inner cylinder to be force free, the viscodiffusive contribution to the stress from the imposed concentration gradient is balanced by viscous contributions from a velocity gradient, leading to the cylinder's rotation. The rotation rate scales linearly with $\tilde{\ell}$, as seen in Fig. 3(d), just as the total current $\tilde{Q}$ does for the chiral generator. In the limit $\tilde{\ell} \to 0$, we obtain $\tilde{\Omega} = \xi(1+\xi)\tilde{\ell}$, where the dimensional rotation rate is $\Omega = \phi\Delta c\tilde{\Omega}/[\xi(\lambda_2 + \lambda_3)L]$. This expression, along with measuring $\Omega$ for a concentration difference $\Delta c$ yields an independent estimate for the coefficient $\phi$.

In summary, the two model systems, i.e. the chiral generator and chiral engine, provide independent estimates for $\psi$ and $\phi$, respectively. In the specific case of passive fluids, where TRS is preserved, the two measured coefficients may be used to experimentally verify the reciprocal relation [19].

### Chiral Bacterial Solutions as Viscodiffusive Fluids

Viscodiffusive fluids are characterized by PS-breaking correlations between a flux and antisymmetric stress, as illustrated in the Green-Kubo relations [10], [11], and [12]. Chiral bacterial solutions represent a natural choice to realize such correlations, as depicted for helical geometries in Fig. 2. Furthermore, the constitutive laws [4] and [21] indicate that such solutions when sheared should exhibit odd diffusive fluxes perpendicular to the shear plane. Recent experiments (23) display such a response when considering solutions of chiral helical immotile bacteria (spirochetes). In particular, Marcos et al. (23) study the migration of spirochetes in a microfluidic channel subjected to a Poiseuille flow (see Fig 4(a) for a schematic of the setup). Upon being injected as a line source at the start of the channel, the spirochetes demonstrate a height-dependent lateral drift that is proportional and perpendicular to the flow's shear rate. Specifically, spirochetes at the top/bottom of the channel migrate to the right/left as they flow down the channel, as shown in Fig. 4(a). Considering spirochete solutions as viscodiffusive fluids, we analyze the experiment by Marcos et al. (23) using the theory presented here. Since the bacteria are *immotile*, we use the constitutive laws relevant for *passive* chiral liquids given by Eqs. [21] and [22], [†] and provide an estimate of the single viscodiffusive coefficient $\psi$.

In analyzing the setup, we adopt a coordinate system identical to Ref. (23) along with the ansatzes $c = c(x, y, z)$, $\mathbf{v} = v_x(x, y, z)\mathbf{e}_x$, and $\mathbf{M} = M_y(x, y, z)\mathbf{e}_y + M_z(x, y, z)\mathbf{e}_z$. In practice, we solve for the angular velocity field $\boldsymbol{\omega}$ rather than the spin angular momentum field $\mathbf{M}$ using the relation $\rho\mathbf{M} = I\boldsymbol{\omega}$. Since the particle moments of inertia of the solute and solvent are not identical, the moment of inertia density $I$ is chosen to be an average

$$I(c) = \frac{m_s c_s d_s^2 + m_c c d_c^2}{4} \quad [25]$$

$$= \frac{\rho d_c^2}{4}\left[\frac{m_c c}{\rho}\left(1 - \frac{d_s^2}{d_c^2}\right) + \frac{d_s^2}{d_c^2}\right], \quad [26]$$

with solvent concentration $c_s$, $m_c$ and $m_s$ solute and solvent masses, and $d_c$ and $d_s$ solute and solvent particle diameters, respectively. The relevant scales in the problem suggest to treat it as an initial value problem in the $x$-dimension and a boundary value problem in the $y$- and $z$-dimensions. The boundary conditions for the $yz$ plane are then those of no-flux ($\mathbf{J} \cdot \mathbf{n} = 0$), no-slip ($\mathbf{v} = \mathbf{0}$), and zero-spin ($\mathbf{M} = \mathbf{0}$). We then solve Eqs. [7]-[9] using the finite element method (37–42)— see Appendix D for full nondimensionalization and numerical details, including a table that details the reported material parameters pertinent to the experiment, leaving the viscodiffusive coefficient as a lone fitting parameter for a choice of solute weight fraction 10%.

As shown in Fig. 4(b), the viscodiffusive constitutive laws capture reasonably the average drift of spirochetes as reported in Ref. (23), with an estimate of the governing viscodiffusive coefficient $\psi \approx 0.005$ $\mu$m. The model also predicts qualitatively the probability density distributions at the top ($\tilde{y} = 0.75$) and bottom ($\tilde{y} = 0.25$) slices as shown in Fig. 4(c); although the widths of the distributions may be overestimated in experiments due to limitations in spatial resolution as well as our model due to the necessity to add numerical diffusion. At the center of the channel, the velocity gradient is zero and thus the solute solely moves through Fickian diffusion. In accordance with Poiseuille flow, the velocity gradient increases in magnitude but differs in sign when moving towards the lower and upper boundaries of the channel. Thus, due to their chirality, the spirochetes demonstrate viscodiffusive fluxes where they migrate to the right/left at the top/bottom half of the channel, as shown in Fig. 4(c,d). In addition, Fig. 4(d) indicates edge flows that circulate the solute along the boundaries in the clockwise direction, a feature observed in other odd transport phenomena (12, 36, 43).

---

[†] In reality, one may require constitutive laws for which the transport tensors are dependent on the bacterial director field. In this work, we interpret the problem from an isotropic perspective and leave the exploration of anisotropic media to future work, wherein all transport tensors must become director dependent.

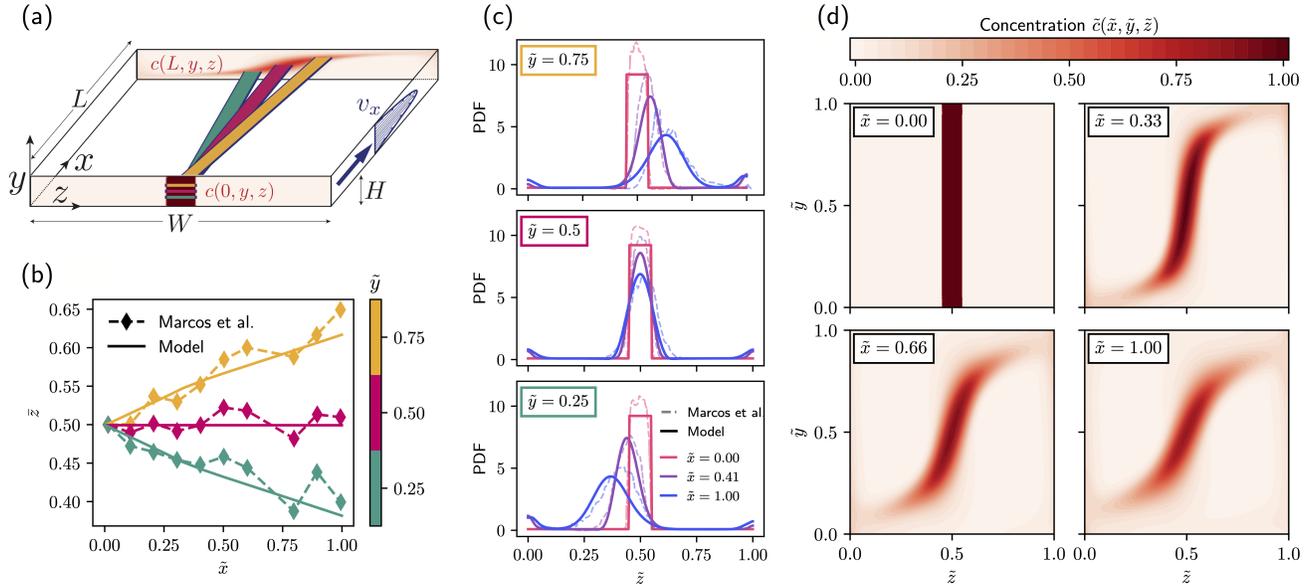

**Fig. 4.** Analysis of emergent viscodiffusive effects in drifting chiral immotile bacteria in Poiseuille flow. **(a)** Schematic adapted from the experimental setup shown in Fig. 2c of (23). We employ an initial condition consisting of a band of bacteria along the $y$ axis, in slight contrast to (23), who inject the bacteria in three discrete locations corresponding to $\tilde{y} \in \{0.25, 0.5, 0.75\}$. **(b)** Average drift $\bar{z}$ in dimensionless units along the channel obtained from model (solid) and compared to the experimental data from (23) (dashed). The results from the model are obtained via numerical simulations using the finite element method (see Appendix D for details). **(c)** Concentration probability density functions (PDF) obtained from the model at three distinct locations along the channel (solid) compared to the data from (23) (dashed). **(d)** Concentration heatmaps at various cross-sections along the channel, where solutes drift in the direction of the vorticity vector.

## Discussion and Conclusions

**Remarks on Curie's Symmetry Principle.** One major assertion of this article is that linear maps of odd rank may exist. In particular, viscodiffusive fluids represent an example of such couplings that emerge under PS-breaking systems, coupling a diffusive flux (rank one) with a velocity gradient (rank two) and a stress (rank two) with a concentration gradient (rank one). Such linear couplings are traditionally neglected as they supposedly violate Curie's symmetry principle (17–22). As a result, some have attributed viscodiffusive transport to be strictly a nonlinear effect (20, 21). However, as also discussed in Appendix A.1 of Ref. (44), the principle merely states, "*(1) when certain causes produce certain effects, the symmetry elements of the causes must be found in the effects produced, and (2) when certain effects reveal a certain asymmetry, this asymmetry must be found in the causes which gave rise to it*" (1). ¶

Since its inception, Curie's principle has been invoked to refute tensor couplings of unalike rank, but in fact such arguments stem from other considerations. De Groot and Mazur (17) and Evans and Morriss (18) demonstrate that for matrices, vectors, and scalars, couplings of unalike rank vanish from isotropic systems that are *also* invariant under inversion (PS-preserving). Finlayson and Scriven (44) state while couplings of unalike rank may exist, couplings between viscous and diffusive transport vanish as they map polar tensors to axial tensors. However, we note that in chiral systems, PS is broken and thus mappings between polar and axial tensors are permitted. In other words, the signs of the viscodiffusive coefficients $\psi$, $\eta$, and $\phi$ simply change upon mirror inversion, as depicted in Fig. 2. Hence, couplings between tensors of different rank may persist, and are not at odds with the Curie principle.

**Conclusions.** In this article we introduce an example of 3D odd transport phenomena by coupling viscous and diffusive transport. In such "odd viscodiffusive fluids," diffusive fluxes arise orthogonally to a velocity gradient, and reciprocal antisymmetric stresses arise from a concentration gradient. By constructing Green-Kubo relations, we show that these fluids must break PS, but need not break TRS, allowing them to be either passive or active. In the former case, TRS unifies viscodiffusive transport under one governing transport coefficient.

Construction of the chiral generator and chiral engine depicted in this work represents a promising direction to independently experimentally measure the viscodiffusive coefficients $\psi$ and $\phi$, which become unified in passive systems. These examples further show the nature of mechanochemical transduction in these viscodiffusive fluids. Identifying chiral bacterial solutions as viscodiffusive fluids, the proposed constitutive laws reproduce observed drift of bacteria orthogonal to shearing planes in microfluidic channels undergoing Poiseuille flow, as studied in Ref. (23). Although we apply our theory to the passive dilute bacterial solutions in Ref. (23), it is also amenable to active systems. In particular, it may be of interest to understand the emergence of viscodiffusive effects in active solutions consisting of motile chiral constituents and the associated rheotaxis and chemotaxis (45, 46). Moreover, one may also formulate an anisotropic viscodiffusive theory that accounts for both interparticle and flow-induced alignment. Lastly, we hope that this work also inspires exploration of other cross-coupled odd transport phenomena in both active and passive systems.

---
¶ Translated from French.

**ACKNOWLEDGMENTS.** We are grateful to Lila Sarfati, Joshua Fernandes, Ahmad Alkadri, and Hyeongjoo Row for stimulating discussions. A.D. is supported by the National Science Foundation Graduate Research Fellowship Program under Grant No. DGE 2146752. K.K.M, and the genesis of this work, is supported by Director, Office of Science, Office of Basic Energy Sciences, of the U.S. Department of Energy under contract No. DEAc02-05CH11231. This research used resources of the National Energy Research Scientific Computing Center (NERSC), a U.S. Department of Energy Office of Science User Facility located at Lawrence Berkeley National Laboratory, using NERSC award BES-ERCAP0023682.

## A. Derivation of Constitutive Laws from Representation Theorems

Here, through the representation theorem for isotropic tensors (see SI-I of Ref. (16)), we derive the constitutive laws presented in the form of Eqs. [4], [5], and [6] in the main text. We first propose linear laws for the diffusive flux $\mathbf{J}$ and stress tensor $\mathbf{T}$, allowing each quantity to depend on the velocity gradient $\boldsymbol{\nabla}\mathbf{v}$, spin field $\mathbf{M}$, and concentration gradient $\boldsymbol{\nabla}c$. We neglect contributions of the spin field gradient $\boldsymbol{\nabla}\mathbf{M}$ on these fluxes. These linear constitutive laws can be written as the following linear maps:

$$\mathbf{J} = \mathbf{j} - \tilde{\psi}^{(3)}\boldsymbol{\nabla}\mathbf{v} - \tilde{\eta}\mathbf{M} - \mathbf{D}\boldsymbol{\nabla}c, \qquad [27]$$

$$\mathbf{T} = -\mathbf{p} + \boldsymbol{\lambda}^{(4)}\boldsymbol{\nabla}\mathbf{v} + \boldsymbol{\gamma}^{(3)}\mathbf{M} - \boldsymbol{\phi}^{(3)}\boldsymbol{\nabla}c, \qquad [28]$$

where the superscripts indicate the rank of the tensors. In the indicial notation, they are

$$J_i = j_i - \tilde{\psi}_{ijk}v_{j,k} - \tilde{\eta}_{ij}M_j - D_{ij}c_{,j}, \qquad [29]$$

$$T_{ij} = -p_{ij} + \lambda_{ijkl}v_{k,l} + \gamma_{ijk}M_k - \phi_{ijk}c_{,k}. \qquad [30]$$

It should be noted that couplings of the form $\tilde{\psi}_{ijk}$ and $\phi_{ijk}$ are often disregarded due to the Curie principle (1), which many have taken to state that tensors of different rank will not couple to each other (17, 44). However, as we shall see representation theorems indicate that such couplings between pairs of tensors of unalike rank do exist. We write a similar constitutive relation for the couple stress tensor $\mathbf{C}$. However, as we neglect the contributions of the spin gradient on $\mathbf{J}$ and $\mathbf{T}$, we accordingly neglect the contributions of concentration gradient, velocity gradient, and spin field on $\mathbf{C}$—if included, these phenomena would be connected through additional reciprocal relations. In this simplified scenario, the constitutive law for the couple stress becomes

$$\mathbf{C} = -\mathbf{p}_C + \boldsymbol{\alpha}^{(4)}\boldsymbol{\nabla}\mathbf{M}, \qquad [31]$$

or in indicial notation,

$$C_{ij} = -p_{Cij} + \alpha_{ijkl}M_{k,l}. \qquad [32]$$

We now invoke the condition of isotropy. For isotropic systems, tensors of rank $n$ in $m$ dimensions can be represented as a linear combination of the Kronecker delta $\delta_{ij}$ and Levi-Civita epsilon $\epsilon_{i_1 i_2 \ldots i_m}$ (16). For an arbitrary tensor $A^{(n)}_{i_1 i_2 \ldots i_n}$ of rank $n$, the following representations can then be written for 3D isotropic systems:

$$A^{(1)}_i = 0, \qquad [33]$$

$$A^{(2)}_{ij} = A^{(2)}\delta_{ij}, \qquad [34]$$

$$A^{(3)}_{ijk} = A^{(3)}\epsilon_{ijk}, \qquad [35]$$

$$A^{(4)}_{ijkl} = A^{(4)}_1 \delta_{ij}\delta_{kl} + A^{(4)}_2 (\delta_{ik}\delta_{jl} + \delta_{il}\delta_{jk} - \frac{2}{3}\delta_{ij}\delta_{kl}) + A^{(4)}_3 (\delta_{ik}\delta_{jl} - \delta_{il}\delta_{jk}). \qquad [36]$$

with $\epsilon_{ijk}$ being the three-dimensional Levi-Civita tensor. These indicate that there exists no isotropic vector, the isotropic rank two tensor is purely diagonal, and the isotropic rank three tensor is purely antisymmetric. Using these representations, the constitutive laws for the three fluxes become

$$J_i = -\tilde{\psi}\epsilon_{ijk}v_{j,k} - \tilde{\eta}M_i - Dc_{,i}, \qquad [37]$$

$$T_{ij} = -p\delta_{ij} + \lambda_1 v_{k,k}\delta_{ij} + \lambda_2(v_{i,j} + v_{j,i} - \frac{2}{3}v_{k,k}\delta_{ij}) + \lambda_3(v_{i,j} - v_{j,i}) + \gamma\epsilon_{ijk}M_k - \phi\epsilon_{ijk}c_{,k}, \qquad [38]$$

$$C_{ij} = -p_C\delta_{ij} + \alpha_1 M_{k,k}\delta_{ij} + \alpha_2(M_{i,j} + M_{j,i} - \frac{2}{3}M_{k,k}\delta_{ij}) + \alpha_3(M_{i,j} - M_{j,i}), \qquad [39]$$

with $\tilde{\psi}$, $\tilde{\eta}$, $D$, $\lambda_1$, $\lambda_2$, $\lambda_3$, $\gamma$, $\phi$, $\alpha_1$, $\alpha_2$, $\alpha_3$ as the corresponding transport coefficients. Transport coefficients may in general depend on field quantities such as the solute concentration $c$. While most transport coefficients will attain a nonzero value for a solute-free system, $\tilde{\psi}$ and $\tilde{\eta}$ will not. As $\mathbf{J} = \mathbf{0}$ when $c = 0$, the coefficients $\tilde{\psi}$ and $\tilde{\eta}$ must be linear with $c$ to the lowest order. Thus, without loss of generality, we define $\tilde{\psi} := \psi c$ and $\tilde{\eta} := \eta c$. In vectorial notation, we then recover Eqs. [4], [5], and [6] of the main text:

$$\mathbf{J} = \psi c \boldsymbol{\nabla} \times \mathbf{v} - \eta c \mathbf{M} - D\boldsymbol{\nabla}c, \qquad [40]$$

$$\mathbf{T} = -p\boldsymbol{\delta} + (\lambda_1 - \frac{2}{3}\lambda_2)(\boldsymbol{\nabla} \cdot \mathbf{v})\boldsymbol{\delta} + \lambda_2(\boldsymbol{\nabla}\mathbf{v} + \boldsymbol{\nabla}\mathbf{v}^\mathsf{T}) + \lambda_3(\boldsymbol{\nabla}\mathbf{v} - \boldsymbol{\nabla}\mathbf{v}^\mathsf{T}) + \gamma\boldsymbol{\epsilon} \cdot \mathbf{M} - \phi\boldsymbol{\epsilon} \cdot \boldsymbol{\nabla}c, \qquad [41]$$

$$\mathbf{C} = -p_C\boldsymbol{\delta} + (\alpha_1 - \frac{2}{3}\alpha_2)(\boldsymbol{\nabla} \cdot \mathbf{M})\boldsymbol{\delta} + \alpha_2(\boldsymbol{\nabla}\mathbf{M} + \boldsymbol{\nabla}\mathbf{M}^\mathsf{T}) + \alpha_3(\boldsymbol{\nabla}\mathbf{M} - \boldsymbol{\nabla}\mathbf{M}^\mathsf{T}). \qquad [42]$$

It should be noted in two-dimensional systems the couple stress tensor becomes a vector and the spin field becomes a scalar. Moreover, in two-dimensional systems, there exists no rank three isotropic tensor, and thus viscodiffusive fluxes corresponding to $\tilde{\psi}, \tilde{\eta}$, and $\phi$ will vanish.

Note, in general, one may consider the non-conjugate couplings of $\mathbf{J}$ and $\mathbf{T}$ with $\boldsymbol{\nabla}\mathbf{M}$ and of $\mathbf{C}$ with $\boldsymbol{\nabla}c$, $\boldsymbol{\nabla}\mathbf{v}$, and $\mathbf{M}$. By disregarding them here, we assume that there exist no correlations of the microscopic fluctuating couple stress $\delta\hat{\mathbf{C}}$ with the microscopic fluctuating diffusive flux $\delta\hat{\mathbf{J}}$ or with the microscopic fluctuating stress $\delta\hat{\mathbf{T}}$. Accounting for these couplings would introduce 9 additional transport coefficients in isotropic systems. In equilibrium systems where time-reversal symmetry and invariance to rigid body rotation is preserved, these 9 coefficients will reduce to 4 coefficients through reciprocal relations. In this work, we explore the simpler model presented in Eqs. [4]-[6] of the main text.

## B. Derivation of Green-Kubo Relations Through the Flux Hypothesis

Here, we present in detail the derivation of Green-Kubo relations for each transport coefficient using the flux hypothesis recently introduced in Ref. (31). Traditionally, one may have been able to derive the Green-Kubo relations using the Onsager regression hypothesis, which is solely based on the balance laws when considering the decay of small perturbations about steady states (2, 3, 16, 30). However, as we shall see, the odd transport phenomena governed by $\psi$, $\eta$, and $\phi$ do not manifest in the linearized forms of the constitutive laws in conjunction with the balance laws. It is for this reason that the flux hypothesis, a generalization of the regression hypothesis at the level of constitutive laws, is necessary to understand the microscopic underpinnings of the viscodiffusive macroscopic transport coefficients.

First, we recapitulate the results of Ref. (31) below. We begin with the constitutive law

$$\mathbf{J} = -\mathbf{Y}\boldsymbol{\nabla}A, \qquad [43]$$

with flux $\mathbf{J}$, driving force $\boldsymbol{\nabla} A$ with conserved quantity $A$, and transport tensor $\mathbf{Y}$. Using the Fourier representation $A^{\mathbf{q}}(t) = \int d\mathbf{x}\, e^{-i\mathbf{q}\cdot\mathbf{x}} A(\mathbf{x}, t)$ with wave vector $\mathbf{q}$, we obtain

$$\mathbf{J}^{\mathbf{q}} = -\mathbf{Y} \cdot i\mathbf{q} A^{\mathbf{q}}. \qquad [44]$$

The flux hypothesis states that microscopic flux fluctuations in the steady state on average obey macroscopic constitutive behavior, i.e.

$$\left\langle \delta \hat{\mathbf{J}}^{\mathbf{q}}(\Delta t) \right\rangle_{\{\delta \hat{A}^{\mathbf{q}}(0) = a^{\mathbf{q}}\}} = -\mathbf{Y} \cdot i\mathbf{q} a^{\mathbf{q}}. \qquad [45]$$

where $\delta \hat{\mathbf{J}}^{\mathbf{q}}$ and $\delta \hat{A}^{\mathbf{q}}$ are the microscopic fluctuations associated with the macroscopic fields $\mathbf{J}^{\mathbf{q}}$ and $A^{\mathbf{q}}$, respectively. The ensemble average $\langle \cdot \rangle$ is conditioned on observing an initial state $\delta \hat{A}^{\mathbf{q}} = a^{\mathbf{q}}$ at time $t=0$. The time $\Delta t$ is chosen such that it is small compared to the macroscopic relaxation time and large compared to the decay of microscopic correlation functions. The balance law, on the other hand,

$$\frac{\partial \delta \hat{A}}{\partial t} = -\boldsymbol{\nabla} \cdot \delta \hat{\mathbf{J}}, \qquad [46]$$

holds exactly on the level of microscopic fluctuations (19) and can be written in the Fourier representation as

$$\frac{\partial \delta \hat{A}^{\mathbf{q}}}{\partial t} = -i\mathbf{q} \cdot \delta \hat{\mathbf{J}}^{\mathbf{q}}. \qquad [47]$$

Using Eqs. [45] and [47] we may recover the Green-Kubo relation in the small wave vector limit $\mathbf{q} \to \mathbf{0}$

$$\mathbf{Y} = \lim_{\mathbf{q} \to \mathbf{0}} \frac{1}{\left\langle \delta \hat{A}^{\mathbf{q}} \delta \hat{A}^{-\mathbf{q}} \right\rangle} \int_0^{\Delta t} dt \left\langle \delta \hat{\mathbf{J}}^{\mathbf{q}}(t) \otimes \delta \hat{\mathbf{J}}^{-\mathbf{q}}(0) \right\rangle, \qquad [48]$$

which captures both the symmetric and antisymmetric parts of the transport matrix $\mathbf{Y}$; see Ref. (31) for additional details on the flux hypothesis and its application to other odd transport phenomena, including odd diffusivity and viscosity in two-dimensional systems.

We now apply the above methodology to the case of viscodiffusive fluids. We consider an aged system that has reached a spatially homogeneous steady state with bulk mass density $\rho_0$ and bulk solute concentration $c_0$. There exists no net bulk velocity $\mathbf{v}_0 = \mathbf{0}$, and the average spin field is $\mathbf{M}_0 = \mathbf{0}$. Now, considering a steady-state that is free of body forces and body torques, we may write evolution equations for the macroscopic perturbations from this steady state as

$$\frac{\partial \delta c}{\partial t} = -c_o \boldsymbol{\nabla} \cdot \delta \mathbf{v} - \boldsymbol{\nabla} \cdot \delta \mathbf{J}, \qquad [49]$$

$$\rho_0 \frac{\partial \delta \mathbf{v}}{\partial t} = \boldsymbol{\nabla} \cdot \delta \mathbf{T}, \qquad [50]$$

$$\rho_0 \frac{\partial \delta \mathbf{M}}{\partial t} = \boldsymbol{\nabla} \cdot \delta \mathbf{C} - \boldsymbol{\epsilon} : \delta \mathbf{T}, \qquad [51]$$

where we have only considered terms up to linear order in the perturbations. For small perturbations about homogeneous steady states, the linearized macroscopic form of the constitutive laws [4]-[6] in the main text are

$$\delta \mathbf{J} = \psi c_0 \boldsymbol{\nabla} \times \delta \mathbf{v} - \eta c_0 \delta \mathbf{M} - D \boldsymbol{\nabla} \delta c, \qquad [52]$$

$$\delta \mathbf{T} = (\lambda_1 - \frac{2}{3}\lambda_2)(\boldsymbol{\nabla} \cdot \delta \mathbf{v})\boldsymbol{\delta} + \lambda_2(\boldsymbol{\nabla}\delta\mathbf{v} + \boldsymbol{\nabla}\delta\mathbf{v}^{\mathsf{T}}) + \lambda_3(\boldsymbol{\nabla}\delta\mathbf{v} - \boldsymbol{\nabla}\delta\mathbf{v}^{\mathsf{T}}) + \gamma \boldsymbol{\epsilon} \cdot \delta\mathbf{M} - \phi \boldsymbol{\epsilon} \cdot \boldsymbol{\nabla}\delta c, \qquad [53]$$

$$\delta \mathbf{C} = (\alpha_1 - \frac{2}{3}\alpha_2)(\boldsymbol{\nabla} \cdot \delta\mathbf{M})\boldsymbol{\delta} + \alpha_2(\boldsymbol{\nabla}\delta\mathbf{M} + \boldsymbol{\nabla}\delta\mathbf{M}^{\mathsf{T}}) + \alpha_3(\boldsymbol{\nabla}\delta\mathbf{M} - \boldsymbol{\nabla}\delta\mathbf{M}^{\mathsf{T}}). \qquad [54]$$

Substituting Eqs. [52]-[54] into [49]-[51] shows that the odd viscodiffusive effects do not enter the linear evolution equations, and therefore are not amenable to analysis using the regression hypothesis.

In what follows we use the flux hypothesis and obtain Green-Kubo relations for all transport coefficients. The system exhibits *microscopic* fluctuations (denoted by hats) $\delta \hat{A}$ from the homogeneous steady state $\left\langle \hat{A} \right\rangle$. For example, $\delta \hat{c}(\mathbf{x}, t) = \hat{c}(\mathbf{x}, t) - c_0$ denotes the microscopic fluctuations in solute concentration. The balance laws [49], [50] and [51] are valid instantaneously for any microscopic realization of the system. To that end, using the Fourier representation and Einstein indicial notation, the evolution equations for microscopic flucuations become

$$\frac{\partial \delta \hat{c}^{\mathbf{q}}}{\partial t} = -c_0 i q_k \delta \hat{v}_k^{\mathbf{q}} - i q_k \delta \hat{J}_k^{\mathbf{q}}, \qquad [55]$$

$$\rho_0 \frac{\partial \delta \hat{v}_i^{\mathbf{q}}}{\partial t} = i q_j \delta \hat{T}_{ij}^{\mathbf{q}}, \qquad [56]$$

$$\rho_0 \frac{\partial \delta \hat{M}_i^{\mathbf{q}}}{\partial t} = i q_j \delta \hat{C}_{ij}^{\mathbf{q}} - \epsilon_{ijk} \delta \hat{T}_{jk}^{\mathbf{q}}. \qquad [57]$$

We now apply the flux hypothesis to the linearized constitutive laws [52]-[54]. In doing so, we make use of the following notation for integrated time correlation functions:

$$\{\hat{A}\hat{B}\} := \frac{1}{V} \lim_{\mathbf{q} \to \mathbf{0}} \int_0^{\Delta t} dt \left\langle \delta \hat{A}^{\mathbf{q}}(t) \delta \hat{B}^{-\mathbf{q}}(0) \right\rangle, \qquad [58]$$

$$\{\hat{A}\hat{B}\}_{\mathbb{I}}^k := \frac{1}{V} \lim_{\mathbf{q} \to \mathbf{0}} \frac{\partial}{\partial i q_k} \int_0^{\Delta t} dt \left\langle \delta \hat{A}^{\mathbf{q}}(t) \delta \hat{B}^{-\mathbf{q}}(0) \right\rangle. \qquad [59]$$

**Green-Kubo Relations Pertaining to Diffusive Flux.** The linearized constitutive law for the diffusive flux given by Eq. [52] can be expressed in the Fourier representation as

$$\delta J_i^{\mathbf{q}} = -\psi \epsilon_{ijk} c_0 \mathrm{i} q_k \delta v_j^{\mathbf{q}} - \eta \delta_{ij} c_0 \delta M_j^{\mathbf{q}} - D \delta_{ij} \mathrm{i} q_j \delta c^{\mathbf{q}}. \qquad [60]$$

Following Ref. (31), the flux hypothesis corresponding to this Fourier form is then expressed as

$$\left\langle \delta \hat{J}_i^{\mathbf{q}}(\Delta t) \right\rangle_{\{\delta \hat{\mathbf{v}}(0), \delta \hat{\mathbf{M}}(0), \delta \hat{c}(0)\}} = -\psi \epsilon_{ijk} c_0 \mathrm{i} q_k \delta \hat{v}_j^{\mathbf{q}}(0) - \eta \delta_{ij} c_0 \delta \hat{M}_j^{\mathbf{q}}(0) - D \delta_{ij} \mathrm{i} q_j \delta \hat{c}^{\mathbf{q}}(0). \qquad [61]$$

Multiplying by $\delta \hat{M}_m^{-\mathbf{q}}(0)$ and then taking a full ensemble average with respect to the steady state yields

$$\left\langle \delta \hat{J}_i^{\mathbf{q}}(\Delta t) \delta \hat{M}_m^{-\mathbf{q}}(0) \right\rangle = -\psi \epsilon_{ijk} c_0 \mathrm{i} q_k \left\langle \delta \hat{v}_j^{\mathbf{q}}(0) \delta \hat{M}_m^{-\mathbf{q}}(0) \right\rangle - \eta \delta_{ij} c_0 \left\langle \delta \hat{M}_j^{\mathbf{q}}(0) \delta \hat{M}_m^{-\mathbf{q}}(0) \right\rangle - D \delta_{ij} \mathrm{i} q_j \left\langle \delta \hat{c}^{\mathbf{q}}(0) \delta \hat{M}_m^{-\mathbf{q}}(0) \right\rangle. \qquad [62]$$

The left hand side correlation function in Eq. [62] can be expanded as

$$\left\langle \delta \hat{J}_i^{\mathbf{q}}(\Delta t) \delta \hat{M}_m^{-\mathbf{q}}(0) \right\rangle = \left\langle \delta \hat{J}_i^{\mathbf{q}}(0) \delta \hat{M}_m^{-\mathbf{q}}(0) \right\rangle + \int_0^{\Delta t} dt \frac{\partial}{\partial t} \left\langle \delta \hat{J}_i^{\mathbf{q}}(t) \delta \hat{M}_m^{-\mathbf{q}}(0) \right\rangle. \qquad [63]$$

Assuming there exist no static correlations between fluxes and field variables and using stationarity, the time derivative may be passed onto the the field variable in the integrand of the second term reducing Eq. [63] to

$$\left\langle \delta \hat{J}_i^{\mathbf{q}}(\Delta t) \delta \hat{M}_m^{-\mathbf{q}}(0) \right\rangle = - \int_0^{\Delta t} dt \left\langle \delta \hat{J}_i^{\mathbf{q}}(t) \frac{\partial \delta \hat{M}_m^{-\mathbf{q}}}{\partial t'}(t') \right\rangle \bigg|_{t'=0}. \qquad [64]$$

Utilizing this expression for the correlation function, Eq. [62] becomes

$$-\int_0^{\Delta t} dt \left\langle \delta \hat{J}_i^{\mathbf{q}}(t) \frac{\partial \delta \hat{M}_m^{-\mathbf{q}}}{\partial t}(t') \right\rangle \bigg|_{t'=0} = -\psi \epsilon_{ijk} c_0 \mathrm{i} q_k \left\langle \delta \hat{v}_j^{\mathbf{q}} \delta \hat{M}_m^{-\mathbf{q}} \right\rangle - \eta \delta_{ij} c_0 \left\langle \delta \hat{M}_j^{\mathbf{q}} \delta \hat{M}_m^{-\mathbf{q}} \right\rangle - D \delta_{ij} \mathrm{i} q_j \left\langle \delta \hat{c}^{\mathbf{q}} \delta \hat{M}_m^{-\mathbf{q}} \right\rangle. \qquad [65]$$

We may now invoke the spin balance [57], which holds for any realization of the system, yielding

$$\frac{\mathrm{i} q_l}{\rho_0} \int_0^{\Delta t} dt \left\langle \delta \hat{J}_i^{\mathbf{q}}(t) \delta \hat{C}_{ml}^{-\mathbf{q}}(0) \right\rangle + \frac{\epsilon_{mln}}{\rho_0} \int_0^{\Delta t} dt \left\langle \delta \hat{J}_i^{\mathbf{q}}(t) \delta \hat{T}_{ln}^{-\mathbf{q}}(0) \right\rangle$$
$$= -\psi \epsilon_{ijk} c_0 \mathrm{i} q_k \left\langle \delta \hat{v}_j^{\mathbf{q}} \delta \hat{M}_m^{-\mathbf{q}} \right\rangle - \eta \delta_{ij} c_0 \left\langle \delta \hat{M}_j^{\mathbf{q}} \delta \hat{M}_m^{-\mathbf{q}} \right\rangle - D \delta_{ij} \mathrm{i} q_j \left\langle \delta \hat{c}^{\mathbf{q}} \delta \hat{M}_m^{-\mathbf{q}} \right\rangle. \qquad [66]$$

In the limit of small wave vectors, we may group terms that are order $\mathcal{O}(q^0)$ to obtain

$$\frac{\epsilon_{mln}}{\rho_0} \int_0^{\Delta t} dt \left\langle \delta \hat{J}_i^{\mathbf{q}}(t) \delta \hat{T}_{ln}^{-\mathbf{q}}(0) \right\rangle = -\eta \delta_{ij} c_0 \left\langle \delta \hat{M}_j^{\mathbf{q}} \delta \hat{M}_m^{-\mathbf{q}} \right\rangle. \qquad [67]$$

As static correlators must be isotropic as argued in Ref. (16), we define $\nu$ as

$$\nu \delta_{ij} := \frac{1}{V} \lim_{\mathbf{q} \to \mathbf{0}} \left\langle \delta \hat{M}_i^{\mathbf{q}} \delta \hat{M}_j^{-\mathbf{q}} \right\rangle. \qquad [68]$$

Upon a switch of indices and the definitions [58] and [68], we obtain the Green Kubo relation for one of the viscodiffusive coefficients $\eta$ as

$$\boxed{\eta = -\frac{\epsilon_{ijk}}{3 \rho_0 c_0 \nu} \{\hat{J}_i \hat{T}_{jk}\}}. \qquad [69]$$

We now seek a Green-Kubo relation for $\psi$. To this end, we multiply Eq. [61] by $\delta v_m^{-\mathbf{q}}(0)$ and take a full ensemble average to obtain

$$\left\langle \delta \hat{J}_i^{\mathbf{q}}(\Delta t) \delta \hat{v}_m^{-\mathbf{q}}(0) \right\rangle = -\psi \epsilon_{ijk} c_0 \mathrm{i} q_k \left\langle \delta \hat{v}_j^{\mathbf{q}}(0) \delta \hat{v}_m^{-\mathbf{q}}(0) \right\rangle - \eta \delta_{ij} c_0 \left\langle \delta \hat{M}_j^{\mathbf{q}}(0) \delta \hat{v}_m^{-\mathbf{q}}(0) \right\rangle - D \delta_{ij} \mathrm{i} q_j \left\langle \delta \hat{c}^{\mathbf{q}}(0) \delta \hat{v}_m^{-\mathbf{q}}(0) \right\rangle. \qquad [70]$$

Following similar steps as done previously, we simplify the correlation function on the left hand side of Eq. [70] to

$$-\int_0^{\Delta t} dt \left\langle \delta \hat{J}_i^{\mathbf{q}}(t) \frac{\partial \delta \hat{v}_m^{-\mathbf{q}}}{\partial t}(t') \right\rangle \bigg|_{t'=0} = -\psi \epsilon_{ijk} c_0 \mathrm{i} q_k \left\langle \delta \hat{v}_j^{\mathbf{q}} \delta \hat{v}_m^{-\mathbf{q}} \right\rangle - \eta \delta_{ij} c_0 \left\langle \delta \hat{M}_j^{\mathbf{q}} \delta \hat{v}_m^{-\mathbf{q}} \right\rangle - D \delta_{ij} \mathrm{i} q_j \left\langle \delta \hat{c}^{\mathbf{q}} \delta \hat{v}_m^{-\mathbf{q}} \right\rangle. \qquad [71]$$

Invoking the linear momentum balance [56] yields

$$\frac{\mathrm{i} q_l}{\rho_0} \int_0^{\Delta t} dt \left\langle \delta \hat{J}_i^{\mathbf{q}}(t) \delta \hat{T}_{ml}^{-\mathbf{q}}(0) \right\rangle = -\psi \epsilon_{ijk} c_0 \mathrm{i} q_k \left\langle \delta \hat{v}_j^{\mathbf{q}} \delta \hat{v}_m^{-\mathbf{q}} \right\rangle - \eta \delta_{ij} c_0 \left\langle \delta \hat{M}_j^{\mathbf{q}} \delta \hat{v}_m^{-\mathbf{q}} \right\rangle - D \delta_{ij} \mathrm{i} q_j \left\langle \delta \hat{c}^{\mathbf{q}} \delta \hat{v}_m^{-\mathbf{q}} \right\rangle. \qquad [72]$$

Once again, we note that, static correlators must be isotropic (16). Thus, the term $\left\langle \delta \hat{c}^{\mathbf{q}} \delta \hat{v}_m^{-\mathbf{q}} \right\rangle$ must be of order $\mathcal{O}(q)$ to lowest order, while $\left\langle \delta \hat{v}_j^{\mathbf{q}} \delta \hat{v}_m^{-\mathbf{q}} \right\rangle$ and $\left\langle \delta \hat{M}_j^{\mathbf{q}} \delta \hat{v}_m^{-\mathbf{q}} \right\rangle$ must be of order $\mathcal{O}(q^0)$ to their lowest order; see Appendix IV of Ref. (16) for details. Note that, even though $\left\langle \delta \hat{M}_j^{\mathbf{q}} \delta \hat{v}_m^{-\mathbf{q}} \right\rangle$ may have a zeroth order wave vector contribution, it must be zero from Eq. [72] for consistency, which leaves us with non-trivial contributions starting from the first order. In that case, defining the static correlators as

$$\mu \delta_{ij} := \frac{1}{V} \lim_{\mathbf{q} \to \mathbf{0}} \left\langle \delta \hat{v}_i^{\mathbf{q}} \delta \hat{v}_j^{-\mathbf{q}} \right\rangle, \qquad [73]$$

$$N_{ij} := \frac{1}{V} \lim_{\mathbf{q} \to \mathbf{0}} \frac{\partial}{\partial \mathrm{i} q_i} \left\langle \delta \hat{M}_j^{\mathbf{q}} \delta \hat{c}^{-\mathbf{q}} \right\rangle, \qquad [74]$$

$$P_{ijk} := \frac{1}{V} \lim_{\mathbf{q} \to \mathbf{0}} \frac{\partial}{\partial \mathrm{i} q_i} \left\langle \delta \hat{M}_j^{\mathbf{q}} \delta \hat{v}_k^{-\mathbf{q}} \right\rangle, \qquad [75]$$

Eq. [72] in the limit of small **q** becomes

$$\frac{\mathrm{i}q_l}{\rho_0}\{\hat{J}_i\hat{T}_{ml}\} = -\psi\epsilon_{ijk}c_0\mathrm{i}q_k\mu\delta_{jm} - \eta\delta_{ij}c_0\mathrm{i}q_l P_{ljm}\,. \quad [76]$$

For arbitrary **q**, we then obtain

$$\psi + \frac{\eta\epsilon_{ijk}P_{ijk}}{6\mu} = -\frac{\epsilon_{ijk}}{6\rho_0 c_0\mu}\{\hat{J}_i\hat{T}_{jk}\}\,. \quad [77]$$

Now defining

$$\tau := \frac{1}{3}\epsilon_{ijk}P_{ijk}\,, \quad [78]$$

and using the Green-Kubo relation for $\eta$ in Eq. [69], we obtain the Green-Kubo relation for $\psi$:

$$\boxed{\psi = -\frac{\epsilon_{ijk}(1-\tau/\nu)}{6\rho_0 c_0\mu}\{\hat{J}_i\hat{T}_{jk}\}}\,. \quad [79]$$

We now proceed to derive the Green-Kubo relation for the diffusion coefficient $D$. Accordingly, we multiply Eq. [61] by $\delta\hat{c}^{-\mathbf{q}}(0)$ and with the full ensemble average we obtain

$$\left\langle\delta\hat{J}_i^{\mathbf{q}}(\Delta t)\delta\hat{c}^{-\mathbf{q}}(0)\right\rangle = -\psi\epsilon_{ijk}c_0\mathrm{i}q_k\left\langle\delta\hat{v}_j^{\mathbf{q}}(0)\delta\hat{c}^{-\mathbf{q}}(0)\right\rangle - \eta\delta_{ij}c_0\left\langle\delta\hat{M}_j^{\mathbf{q}}(0)\delta\hat{c}^{-\mathbf{q}}(0)\right\rangle - D\delta_{ij}\mathrm{i}q_j\left\langle\delta\hat{c}^{\mathbf{q}}(0)\delta\hat{c}^{-\mathbf{q}}(0)\right\rangle\,. \quad [80]$$

Following similar steps as with the previous Green-Kubo relations, we simplify the correlation function on the left hand of side of Eq. [80] to

$$-\int_0^{\Delta t} dt\left\langle\delta\hat{J}_i^{\mathbf{q}}(t)\frac{\partial\delta\hat{c}^{-\mathbf{q}}}{\partial t}(t')\right\rangle\bigg|_{t'=0} = -\psi\epsilon_{ijk}c_0\mathrm{i}q_k\left\langle\delta\hat{v}_j^{\mathbf{q}}\delta\hat{c}^{-\mathbf{q}}\right\rangle - \eta\delta_{ij}c_0\left\langle\delta\hat{M}_j^{\mathbf{q}}\delta\hat{c}^{-\mathbf{q}}\right\rangle - D\delta_{ij}\mathrm{i}q_j\left\langle\delta\hat{c}^{\mathbf{q}}\delta\hat{c}^{-\mathbf{q}}\right\rangle\,. \quad [81]$$

With the concentration balance [55] we obtain

$$-c_0\mathrm{i}q_l\int_0^{\Delta t}dt\left\langle\delta\hat{J}_i^{\mathbf{q}}(t)\delta\hat{v}_l^{-\mathbf{q}}(0)\right\rangle - \mathrm{i}q_l\int_0^{\Delta t}dt\left\langle\delta\hat{J}_i^{\mathbf{q}}(t)\delta\hat{J}_l^{-\mathbf{q}}(0)\right\rangle = -\psi\epsilon_{ijk}c_0\mathrm{i}q_k\left\langle\delta\hat{v}_j^{\mathbf{q}}\delta\hat{c}^{-\mathbf{q}}\right\rangle - \eta\delta_{ij}c_0\left\langle\delta\hat{M}_j^{\mathbf{q}}\delta\hat{c}^{-\mathbf{q}}\right\rangle - D\delta_{ij}\mathrm{i}q_j\left\langle\delta\hat{c}^{\mathbf{q}}\delta\hat{c}^{-\mathbf{q}}\right\rangle\,. \quad [82]$$

Defining the concentration autocorrelator as

$$s := \frac{1}{c_0 V}\lim_{\mathbf{q}\to\mathbf{0}}\left\langle\delta\hat{c}^{\mathbf{q}}\delta\hat{c}^{-\mathbf{q}}\right\rangle\,. \quad [83]$$

and grouping terms of order $\mathcal{O}(\mathbf{q})$, Eq. [82] in the small wave vector limit becomes

$$c_0\mathrm{i}q_l\{\hat{J}_i\hat{v}_l\} + \mathrm{i}q_l\{\hat{J}_i\hat{J}_l\} = \eta\delta_{ij}c_0\mathrm{i}q_l N_{lj} + D\delta_{ij}\mathrm{i}q_j sc_0\,. \quad [84]$$

Given arbitrary **q**,

$$D = \frac{1}{3sc_0}\left[c_0\{\hat{J}_i\hat{v}_i\} + \{\hat{J}_i\hat{J}_i\}\right] - \frac{\eta N_{ii}}{3s}\,. \quad [85]$$

With the Green-Kubo relation for $\eta$ in Eq. [69], we obtain

$$D = \frac{1}{3sc_0}\left[c_0\{\hat{J}_i\hat{v}_i\} + \{\hat{J}_i\hat{J}_i\}\right] + \frac{\epsilon_{ijk}N_{ll}}{9sc_0\rho_0\nu}\{\hat{J}_i\hat{T}_{jk}\}\,. \quad [86]$$

Note one would expect in most systems that $\{\hat{J}_i\hat{v}_i\} = 0$, which then simplifies Eq. [86] to

$$\boxed{D = \frac{1}{3sc_0}\{\hat{J}_i\hat{J}_i\} + \frac{\epsilon_{ijk}N_{ll}}{9sc_0\rho_0\nu}\{\hat{J}_i\hat{T}_{jk}\}}\,. \quad [87]$$

Lastly, for systems with no static correlations between spin and concentration fluctuations, $N_{ll} = 0$ and the Green-Kubo relation for the diffusion coefficient becomes

$$\boxed{D = \frac{1}{3sc_0}\{\hat{J}_i\hat{J}_i\}}\,. \quad [88]$$

**Green-Kubo Relations Pertaining to the Stress Tensor.** We now proceed to derive the Green-Kubo relations for the transport coefficients associated with the constitutive law for the stress and the linear momentum balance. To that end, the linearized constitutive law in Eq. [53] can be expressed in Fourier representation as

$$\delta T_{ij}^{\mathbf{q}} = \sum_{\alpha=1}^{3}\lambda_\alpha s_{ijkl}^\alpha \mathrm{i}q_l\delta v_k^{\mathbf{q}} + \gamma\epsilon_{ijk}\delta M_k^{\mathbf{q}} - \phi\epsilon_{ijk}\mathrm{i}q_k\delta c^{\mathbf{q}}\,, \quad [89]$$

which then yields the following form for the behavior of fluctuations from the flux hypothesis:

$$\left\langle\delta\hat{T}_{ij}^{\mathbf{q}}(\Delta t)\right\rangle_{\{\delta\hat{\mathbf{v}}(0),\delta\hat{\mathbf{M}}(0),\delta\hat{c}(0)\}} = \sum_{\alpha=1}^{3}\lambda_\alpha s_{ijkl}^\alpha \mathrm{i}q_l\delta\hat{v}_k^{\mathbf{q}}(0) + \gamma\epsilon_{ijk}\delta\hat{M}_k^{\mathbf{q}}(0) - \phi\epsilon_{ijk}\mathrm{i}q_k\delta\hat{c}^{\mathbf{q}}(0)\,. \quad [90]$$

Note that, in writing the above equations, we have used the notation for the viscosity tensor $\lambda_{ijkl} = \sum_{\alpha=1}^{3}\lambda_\alpha s_{ijkl}^\alpha$ using the three isotropic rank four tensors $\{s_{ijkl}^\alpha\}$ given in Eq. [36]. As before, multiplying Eq. [90] by $\delta\hat{M}_m^{-\mathbf{q}}(0)$ and then performing a full ensemble average yields

$$\left\langle\delta\hat{T}_{ij}^{\mathbf{q}}(\Delta t)\delta\hat{M}_m^{-\mathbf{q}}(0)\right\rangle = \sum_{\alpha=1}^{3}\lambda_\alpha s_{ijkl}^\alpha \mathrm{i}q_l\left\langle\delta\hat{v}_k^{\mathbf{q}}(0)\delta\hat{M}_m^{-\mathbf{q}}(0)\right\rangle + \gamma\epsilon_{ijk}\left\langle\delta\hat{M}_k^{\mathbf{q}}(0)\delta\hat{M}_m^{-\mathbf{q}}(0)\right\rangle - \phi\epsilon_{ijk}\mathrm{i}q_k\left\langle\delta\hat{c}^{\mathbf{q}}(0)\delta\hat{M}_m^{-\mathbf{q}}(0)\right\rangle\,. \quad [91]$$

The left hand side correlation function in Eq. [91] can be expanded as

$$\left\langle \delta\hat{T}_{ij}^{\mathbf{q}}(\Delta t)\delta\hat{M}_m^{-\mathbf{q}}(0)\right\rangle = \left\langle \delta\hat{T}_{ij}^{\mathbf{q}}(0)\delta\hat{M}_m^{-\mathbf{q}}(0)\right\rangle + \int_0^{\Delta t} dt \frac{\partial}{\partial t}\left\langle \delta\hat{T}_{ij}^{\mathbf{q}}(t)\delta\hat{M}_m^{-\mathbf{q}}(0)\right\rangle. \qquad [92]$$

As before, assuming there exist no static correlations between fluxes and field variables, Eq. [92] becomes

$$\left\langle \delta\hat{T}_{ij}^{\mathbf{q}}(\Delta t)\delta\hat{M}_m^{-\mathbf{q}}(0)\right\rangle = -\int_0^{\Delta t} dt \left\langle \delta\hat{T}_{ij}^{\mathbf{q}}(t)\frac{\partial\delta\hat{M}_m^{-\mathbf{q}}}{\partial t}(t')\right\rangle\bigg|_{t'=0}. \qquad [93]$$

which reduces Eq. [91] to

$$-\int_0^{\Delta t} dt \left\langle \delta\hat{T}_{ij}^{\mathbf{q}}(t)\frac{\partial\delta\hat{M}_m^{-\mathbf{q}}}{\partial t}(t')\right\rangle\bigg|_{t'=0} = \sum_{\alpha=1}^3 \lambda_\alpha s_{ijkl}^\alpha \mathrm{i}q_l \left\langle\delta\hat{v}_k^{\mathbf{q}}(0)\delta\hat{M}_m^{-\mathbf{q}}(0)\right\rangle + \gamma\epsilon_{ijk}\left\langle\delta\hat{M}_k^{\mathbf{q}}(0)\delta\hat{M}_m^{-\mathbf{q}}(0)\right\rangle - \phi\epsilon_{ijk}\mathrm{i}q_k\left\langle\delta\hat{c}^{\mathbf{q}}(0)\delta\hat{M}_m^{-\mathbf{q}}(0)\right\rangle. \qquad [94]$$

With the instantaneous spin balance [57], Eq. [94] becomes

$$\frac{\mathrm{i}q_l}{\rho_0}\int_0^{\Delta t} dt \left\langle \delta\hat{T}_{ij}^{\mathbf{q}}(t)\delta\hat{C}_{ml}^{-\mathbf{q}}(0)\right\rangle + \frac{\epsilon_{mln}}{\rho_0}\int_0^{\Delta t} dt \left\langle \delta\hat{T}_{ij}^{\mathbf{q}}(t)\delta\hat{T}_{ln}^{-\mathbf{q}}(0)\right\rangle$$
$$= \sum_{\alpha=1}^3 \lambda_\alpha s_{ijkl}^\alpha \mathrm{i}q_l \left\langle\delta\hat{v}_k^{\mathbf{q}}(0)\delta\hat{M}_m^{-\mathbf{q}}(0)\right\rangle + \gamma\epsilon_{ijk}\left\langle\delta\hat{M}_k^{\mathbf{q}}(0)\delta\hat{M}_m^{-\mathbf{q}}(0)\right\rangle - \phi\epsilon_{ijk}\mathrm{i}q_k\left\langle\delta\hat{c}^{\mathbf{q}}(0)\delta\hat{M}_m^{-\mathbf{q}}(0)\right\rangle. \qquad [95]$$

In the limit of small wave vectors, we see

$$\frac{\epsilon_{mln}}{\rho_0}\int_0^{\Delta t} dt \left\langle \delta\hat{T}_{ij}^{\mathbf{q}}(t)\delta\hat{T}_{ln}^{-\mathbf{q}}(0)\right\rangle = \gamma\epsilon_{ijk}\left\langle\delta\hat{M}_k^{\mathbf{q}}\delta\hat{M}_m^{-\mathbf{q}}\right\rangle. \qquad [96]$$

Upon using the definition [68], we obtain the following Green-Kubo relation for $\gamma$:

$$\boxed{\gamma = \frac{\delta_{ik}\delta_{jl} - \delta_{il}\delta_{jk}}{6\rho_0\nu}\{\hat{T}_{ij}\hat{T}_{kl}\}}. \qquad [97]$$

We move onto deriving the Green-Kubo relations for the viscosities $\lambda_{ijkl}$. To this end, multiplying Eq. [90] with $\delta v_m^{-\mathbf{q}}(0)$ and taking a full ensemble average yields

$$\left\langle \delta\hat{T}_{ij}^{\mathbf{q}}(\Delta t)\delta\hat{v}_m^{-\mathbf{q}}(0)\right\rangle = \sum_{\alpha=1}^3 \lambda_\alpha s_{ijkl}^\alpha \mathrm{i}q_l\left\langle\delta\hat{v}_k^{\mathbf{q}}(0)\delta\hat{v}_m^{-\mathbf{q}}(0)\right\rangle + \gamma\epsilon_{ijk}\left\langle\delta\hat{M}_k^{\mathbf{q}}(0)\delta\hat{v}_m^{-\mathbf{q}}(0)\right\rangle - \phi\epsilon_{ijk}\mathrm{i}q_k\left\langle\delta\hat{c}^{\mathbf{q}}(0)\delta\hat{v}_m^{-\mathbf{q}}(0)\right\rangle. \qquad [98]$$

Following similar steps as done previously, we simplify the correlation function on the left hand side of Eq. [98] to

$$-\int_0^{\Delta t} dt \left\langle \delta\hat{T}_{ij}^{\mathbf{q}}(t)\frac{\partial\delta\hat{v}_m^{-\mathbf{q}}}{\partial t}(t')\right\rangle\bigg|_{t'=0} = \sum_{\alpha=1}^3 \lambda_\alpha s_{ijkl}^\alpha \mathrm{i}q_l \left\langle\delta\hat{v}_k^{\mathbf{q}}\delta\hat{v}_m^{-\mathbf{q}}\right\rangle + \gamma\epsilon_{ijk}\left\langle\delta\hat{M}_k^{\mathbf{q}}\delta\hat{v}_m^{-\mathbf{q}}\right\rangle - \phi\epsilon_{ijk}\mathrm{i}q_k\left\langle\delta\hat{c}^{\mathbf{q}}\delta\hat{v}_m^{-\mathbf{q}}\right\rangle, \qquad [99]$$

which along with the linear momentum balance [56] yields

$$\frac{\mathrm{i}q_l}{\rho_0}\int_0^{\Delta t} dt \left\langle \delta\hat{T}_{ij}^{\mathbf{q}}(t)\delta\hat{T}_{ml}^{-\mathbf{q}}(0)\right\rangle = \sum_{\alpha=1}^3 \lambda_\alpha s_{ijkl}^\alpha \mathrm{i}q_l \left\langle\delta\hat{v}_k^{\mathbf{q}}\delta\hat{v}_m^{-\mathbf{q}}\right\rangle + \gamma\epsilon_{ijk}\left\langle\delta\hat{M}_k^{\mathbf{q}}\delta\hat{v}_m^{-\mathbf{q}}\right\rangle - \phi\epsilon_{ijk}\mathrm{i}q_k\left\langle\delta\hat{c}^{\mathbf{q}}\delta\hat{v}_m^{-\mathbf{q}}\right\rangle. \qquad [100]$$

Using the definitions [73] and [75] and collecting terms of order $\mathcal{O}(\mathbf{q})$, we obtain

$$\frac{\mathrm{i}q_l}{\rho_0}\{\hat{T}_{ij}\hat{T}_{ml}\} = \sum_{\alpha=1}^3 \lambda_\alpha s_{ijkl}^\alpha \mathrm{i}q_l\mu\delta_{km} + \gamma_{ijk}\mathrm{i}q_l P_{lkm}, \qquad [101]$$

which for arbitrary $\mathbf{q}$ yields

$$\sum_{\alpha=1}^3 \lambda_\alpha s_{ijkl}^\alpha + \frac{P_{lmk}\gamma_{ijm}}{\mu} = \frac{1}{\rho_0\mu}\{\hat{T}_{ij}\hat{T}_{kl}\}. \qquad [102]$$

Finally, using the Green-Kubo relation for $\gamma$ in Eq. [97], and the representations for the isotropic tensors $s_{ijkl}^\alpha$, we obtain the following Green-Kubo relation for $\lambda_1$, $\lambda_2$, and $\lambda_3$:

$$\boxed{\lambda_1 = \frac{\delta_{ij}\delta_{kl}}{9\rho_0\mu}\{\hat{T}_{ij}\hat{T}_{kl}\}} \qquad [103]$$

$$\boxed{\lambda_2 = \frac{(\delta_{ik}\delta_{jl} + \delta_{il}\delta_{jk} - \frac{2}{3}\delta_{ij}\delta_{kl})}{20\rho_0\mu}\{\hat{T}_{ij}\hat{T}_{kl}\}} \qquad [104]$$

$$\boxed{\lambda_3 = \frac{(\delta_{ik}\delta_{jl} - \delta_{il}\delta_{jk})(1-\tau/\nu)}{12\rho_0\mu}\{\hat{T}_{ij}\hat{T}_{kl}\}}. \qquad [105]$$

We now seek a Green-Kubo relation for the viscodiffusive coefficient $\phi$. Accordingly, we multiply Eq. [90] by $\delta\hat{c}^{-\mathbf{q}}(0)$ and obtain

$$\left\langle \delta\hat{T}_{ij}^{\mathbf{q}}(\Delta t)\delta\hat{c}^{-\mathbf{q}}(0)\right\rangle = \sum_{\alpha=1}^{3} \lambda_\alpha s_{ijkl}^{\alpha} \mathrm{i}q_l \left\langle \delta\hat{v}_k^{\mathbf{q}}(0)\delta\hat{c}^{-\mathbf{q}}(0)\right\rangle + \gamma\epsilon_{ijk}\left\langle \delta\hat{M}_k^{\mathbf{q}}(0)\delta\hat{c}^{-\mathbf{q}}(0)\right\rangle - \phi\epsilon_{ijk}\mathrm{i}q_k\left\langle \delta\hat{c}^{\mathbf{q}}(0)\delta\hat{c}^{-\mathbf{q}}(0)\right\rangle , \qquad [106]$$

wherein the time-correlation function can be rewritten to yield

$$-\int_0^{\Delta t} dt \left\langle \delta\hat{T}_{ij}^{\mathbf{q}}(t)\frac{\partial\delta\hat{c}^{-\mathbf{q}}}{\partial t}(t')\right\rangle \bigg|_{t'=0} = \sum_{\alpha=1}^{3} \lambda_\alpha s_{ijkl}^{\alpha}\mathrm{i}q_l \left\langle \delta\hat{v}_k^{\mathbf{q}}\delta\hat{c}^{-\mathbf{q}}\right\rangle + \gamma\epsilon ijk\left\langle \delta\hat{M}_k^{\mathbf{q}}\delta\hat{c}^{-\mathbf{q}}\right\rangle - \phi\epsilon_{ijk}\mathrm{i}q_k\left\langle \delta\hat{c}^{\mathbf{q}}\delta\hat{c}^{-\mathbf{q}}\right\rangle . \qquad [107]$$

With the concentration balance [55], we have

$$-c_0\mathrm{i}q_l \int_0^{\Delta t} dt \left\langle \delta\hat{T}_{ij}^{\mathbf{q}}(t)\delta\hat{v}_l^{-\mathbf{q}}(0)\right\rangle - \mathrm{i}q_l \int_0^{\Delta t} dt \left\langle \delta\hat{T}_{ij}^{\mathbf{q}}(t)\delta\hat{J}_l^{-\mathbf{q}}(0)\right\rangle$$
$$= \sum_{\alpha=1}^{3} \lambda_\alpha s_{ijkl}^{\alpha}\mathrm{i}q_l\left\langle \delta\hat{v}_k^{\mathbf{q}}\delta\hat{c}^{-\mathbf{q}}\right\rangle + \gamma\epsilon_{ijk}\left\langle \delta\hat{M}_k^{\mathbf{q}}\delta\hat{c}^{-\mathbf{q}}\right\rangle - \phi\epsilon_{ijk}\mathrm{i}q_k\left\langle \delta\hat{c}^{\mathbf{q}}\delta\hat{c}^{-\mathbf{q}}\right\rangle . \qquad [108]$$

We now group terms of order $\mathcal{O}(\mathbf{q})$ from the leading order contributions of the static correlators. With prior definitions for the static correlators, we therefore obtain

$$c_0\mathrm{i}q_l\{\hat{T}_{ij}\hat{v}_l\} + \mathrm{i}q_l\{\hat{T}_{ij}\hat{J}_l\} = -\gamma\epsilon_{ijk}\mathrm{i}q_l N_{lk} + \phi\epsilon_{ijk}\mathrm{i}q_k sc_0 , \qquad [109]$$

which, for arbitrary $\mathbf{q}$, leads to

$$\phi - \frac{1}{3sc_0}\gamma N_{kk} = \frac{\epsilon_{ijk}}{6sc_0}\left[ c_0\{\hat{T}_{ij}\hat{v}_k\} + \{\hat{T}_{ij}\hat{J}_k\}\right] . \qquad [110]$$

Using the Green-Kubo relation for $\gamma$ [97], we obtain the Green-Kubo relation for the viscodiffusive coefficient

$$\boxed{\phi = \frac{\epsilon_{ijk}}{6sc_0}\left(\{\hat{T}_{ij}\hat{J}_k\} + c_0\{\hat{T}_{ij}\hat{v}_k\}\right) + \frac{N_{mm}(\delta_{ik}\delta_{jl} - \delta_{il}\delta_{jk})}{18\rho_0\nu sc_0}\{\hat{T}_{ij}\hat{T}_{kl}\}} . \qquad [111]$$

For most systems, we would expect no correlations of velocity and stress, and so the above relation reduces to

$$\boxed{\phi = \frac{\epsilon_{ijk}}{6sc_0}\{\hat{T}_{ij}\hat{J}_k\} + \frac{N_{mm}(\delta_{ik}\delta_{jl} - \delta_{il}\delta_{jk})}{18\rho_0\nu sc_0}\{\hat{T}_{ij}\hat{T}_{kl}\}} . \qquad [112]$$

Moreover, we may assume that first-order static correlations of the spin and concentration fields are negligible, yielding a simplified Green-Kubo relation relating the viscodiffusive coefficient to correlations between the antisymmetric stresses and diffusive fluxes as

$$\boxed{\phi = \frac{\epsilon_{ijk}}{6sc_0}\{\hat{T}_{ij}\hat{J}_k\}} . \qquad [113]$$

**Green-Kubo Relations Pertaining to the Couple Stress Tensor.** Lastly, we derive Green-Kubo relations for the spin viscosities $\alpha_1$, $\alpha_2$, and $\alpha_3$, which relate spin gradients to the couple stress tensor. The associated linearized constitutive law [54] may be expressed in the Fourier representation as

$$\delta C_{ij}^{\mathbf{q}} = \sum_{\beta=1}^{3} \alpha_\beta s_{ijkl}^{\beta}\mathrm{i}q_l\delta M_k^{\mathbf{q}} , \qquad [114]$$

with the spin viscosity tensor $\alpha_{ijkl} = \sum_{\beta=1}^{3} \alpha_\beta s_{ijkl}^{\beta}$, $\beta \in \{1,2,3\}$ using the rank-four isotropic basis tensors $\{s_{ijkl}^{\beta}\}$ in Eq. [36]. The corresponding flux hypothesis is

$$\left\langle \delta\hat{C}_{ij}^{\mathbf{q}}(\Delta t)\right\rangle_{\{\delta\hat{\mathbf{M}}(0)\}} = \sum_{\beta=1}^{3} \alpha_\beta s_{ijkl}^{\beta}\mathrm{i}q_l\delta\hat{M}_k^{\mathbf{q}}(0) , \qquad [115]$$

Multiplying by $\delta\hat{M}_m^{-\mathbf{q}}(0)$ and following the same steps as before we obtain

$$-\int_0^{\Delta t} dt \left\langle \delta\hat{C}_{ij}^{\mathbf{q}}(t)\frac{\partial\delta\hat{M}_m^{-\mathbf{q}}}{\partial t}(t')\right\rangle \bigg|_{t'=0} = \sum_{\beta=1}^{3} \alpha_\beta s_{ijkl}^{\beta}\mathrm{i}q_l\left\langle \delta\hat{M}_k^{\mathbf{q}}\delta\hat{M}_m^{-\mathbf{q}}\right\rangle . \qquad [116]$$

With the instantaneous spin balance [57],

$$\frac{\mathrm{i}q_l}{\rho_0}\int_0^{\Delta t} dt \left\langle \delta\hat{C}_{ij}^{\mathbf{q}}(t)\delta\hat{C}_{ml}^{-\mathbf{q}}(0)\right\rangle + \frac{\epsilon_{mln}}{\rho_0}\int_0^{\Delta t} dt \left\langle \delta\hat{C}_{ij}^{\mathbf{q}}(t)\delta\hat{T}_{ln}^{-\mathbf{q}}(0)\right\rangle = \sum_{\beta=1}^{3} \alpha_\beta s_{ijkl}^{\beta}\mathrm{i}q_l\left\langle \delta\hat{M}_k^{\mathbf{q}}\delta\hat{M}_m^{-\mathbf{q}}\right\rangle . \qquad [117]$$

Using the definitions [58], [59], and [68], we group terms of order $\mathcal{O}(\mathbf{q})$:

$$\frac{\mathrm{i}q_l}{\rho_0}\left[\{\hat{C}_{ij}\hat{C}_{ml}\} + \epsilon_{mkn}\{\hat{C}_{ij}\hat{T}_{kn}\}_{\mathbb{I}}^l\right] = \sum_{\beta=1}^{3} \alpha_\beta s_{ijkl}^{\beta}\mathrm{i}q_l\nu\delta_{km} . \qquad [118]$$

For arbitrary **q** we get

$$\sum_{\beta=1}^{3} \alpha_\beta s^\beta_{ijkl} = \frac{1}{\rho_0 \nu} \left[ \{\hat{C}_{ij}\hat{C}_{kl}\} + \epsilon_{kmn}\{\hat{C}_{ij}\hat{T}_{mn}\}^l_\mathbb{I} \right]. \qquad [119]$$

Finally, by contracting with the basis tensors in [36], we isolate the Green-Kubo relations for each of the three spin viscosities:

$$\boxed{\alpha_1 = \frac{\delta_{ij}\delta_{kl}}{9\rho_0 \nu} \left[ \{\hat{C}_{ij}\hat{C}_{kl}\} + \epsilon_{kmn}\{\hat{C}_{ij}\hat{T}_{mn}\}^l_\mathbb{I} \right]} \qquad [120]$$

$$\boxed{\alpha_2 = \frac{(\delta_{ik}\delta_{jl} + \delta_{il}\delta_{jk} - \frac{2}{3}\delta_{ij}\delta_{kl})}{20\rho_0 \nu} \left[ \{\hat{C}_{ij}\hat{C}_{kl}\} + \epsilon_{kmn}\{\hat{C}_{ij}\hat{T}_{mn}\}^l_\mathbb{I} \right]} \qquad [121]$$

$$\boxed{\alpha_3 = \frac{(\delta_{ik}\delta_{jl} - \delta_{il}\delta_{jk})}{12\rho_0 \nu} \left[ \{\hat{C}_{ij}\hat{C}_{kl}\} + \epsilon_{kmn}\{\hat{C}_{ij}\hat{T}_{mn}\}^l_\mathbb{I} \right]}. \qquad [122]$$

**Summary of Green-Kubo Relations.** In what follows, we summarize the Green-Kubo relations for all transport coefficients in Eqs. [4]-[6] in the main text. Here, we have assumed there exist no internal body forces or internal body torques, no static spin-concentration correlations ($N_{ii} = 0$), no flux-velocity correlations ($\{\hat{J}_i \hat{v}_i\} = 0$), no stress-velocity correlations ($\epsilon_{ijk}\{\hat{T}_{ij}\hat{v}_k\} = 0$), and that isotropic conditions are valid. The Green-Kubo relations for the transport coefficients derived before reduce to the following simplified forms:

$$D = \frac{1}{3sc_0}\{\hat{J}_k \hat{J}_k\}, \qquad [123]$$

$$\psi = -\frac{\epsilon_{ijk}(1 - \tau/\nu)}{6\rho_0 c_0 \mu}\{\hat{J}_i \hat{T}_{jk}\}, \qquad [124]$$

$$\eta = -\frac{\epsilon_{ijk}}{3\rho_0 c_0 \nu}\{\hat{J}_i \hat{T}_{jk}\}, \qquad [125]$$

$$\phi = \frac{\epsilon_{ijk}}{6sc_0}\{\hat{T}_{ij}\hat{J}_k\}, \qquad [126]$$

$$\lambda_1 = \frac{\delta_{ij}\delta_{kl}}{9\rho_0 \mu}\{\hat{T}_{ij}\hat{T}_{kl}\}, \qquad [127]$$

$$\lambda_2 = \frac{(\delta_{ik}\delta_{jl} + \delta_{il}\delta_{jk} - \frac{2}{3}\delta_{ij}\delta_{kl})}{20\rho_0 \mu}\{\hat{T}_{ij}\hat{T}_{kl}\}, \qquad [128]$$

$$\lambda_3 = \frac{(\delta_{ik}\delta_{jl} - \delta_{il}\delta_{jk})(1 - \tau/\nu)}{12\rho_0 \mu}\{\hat{T}_{ij}\hat{T}_{kl}\}, \qquad [129]$$

$$\gamma = \frac{(\delta_{ik}\delta_{jl} - \delta_{il}\delta_{jk})}{6\rho_0 \nu}\{\hat{T}_{ij}\hat{T}_{kl}\}, \qquad [130]$$

$$\alpha_1 = \frac{\delta_{ij}\delta_{kl}}{9\rho_0 \nu} \left[ \{\hat{C}_{ij}\hat{C}_{kl}\} + \epsilon_{kmn}\{\hat{C}_{ij}\hat{T}_{mn}\}^l_\mathbb{I} \right], \qquad [131]$$

$$\alpha_2 = \frac{(\delta_{ik}\delta_{jl} + \delta_{il}\delta_{jk} - \frac{2}{3}\delta_{ij}\delta_{kl})}{20\rho_0 \nu} \left[ \{\hat{C}_{ij}\hat{C}_{kl}\} + \epsilon_{kmn}\{\hat{C}_{ij}\hat{T}_{mn}\}^l_\mathbb{I} \right], \qquad [132]$$

$$\alpha_3 = \frac{(\delta_{ik}\delta_{jl} - \delta_{il}\delta_{jk})}{12\rho_0 \nu} \left[ \{\hat{C}_{ij}\hat{C}_{kl}\} + \epsilon_{kmn}\{\hat{C}_{ij}\hat{T}_{mn}\}^l_\mathbb{I} \right]. \qquad [133]$$

## C. Chiral Generator and Chiral Engine: Analytical Details

**Chiral Generator.** Here, we present the analytical solution for the chiral generator, corresponding to the setup in Fig. 3(a). We assume the ansatzes, $c = c_0$, $\mathbf{v} = v_\theta(r)\mathbf{e}_\theta$, and $\mathbf{M} = M_z(r)\mathbf{e}_z$, and defining the following nondimensional quantities

$$\tilde{r} := r/R_{\rm o}, \qquad [134]$$

$$\xi := R_{\rm i}/R_{\rm o}, \qquad [135]$$

$$\tilde{v}_\theta := \frac{v_\theta}{R_i \Omega}, \qquad [136]$$

$$\tilde{M}_z := \frac{M_z}{\lambda_3 \xi \Omega/\gamma}, \qquad [137]$$

$$\tilde{J}_z := \frac{J_z}{\psi c_0 \xi \Omega}, \qquad [138]$$

$$\tilde{T}^{\rm A}_{\theta r} := \frac{T_{\theta r} - T_{r\theta}}{\lambda_3 \xi \Omega}, \qquad [139]$$

Eqs. [8] and [9] from the main text become

$$0 = \frac{d}{d\tilde{r}}\left[\frac{1}{\tilde{r}}\frac{d}{d\tilde{r}}(\tilde{r}\tilde{v}_\theta)\right] - \tilde{\lambda}_3 \frac{d\tilde{M}_z}{d\tilde{r}}, \qquad [140]$$

$$0 = \frac{\rm Sp}{\tilde{r}}\frac{d}{d\tilde{r}}\left(\tilde{r}\frac{d\tilde{M}_z}{d\tilde{r}}\right) + \frac{1}{\tilde{r}}\frac{d}{d\tilde{r}}(\tilde{r}\tilde{v}_\theta) - \tilde{M}_z, \qquad [141]$$

where the following dimensionless groups have been defined:

$$\tilde{\lambda}_3 := \frac{\lambda_3}{\lambda_2 + \lambda_3}, \qquad [142]$$

$$\mathrm{Sp} := \frac{\alpha_2 + \alpha_3}{2\gamma R_\mathrm{o}^2}. \qquad [143]$$

The group $0 < \tilde{\lambda}_3 < 1$ represents the rotational viscosity relative to the sum of the shear and rotational viscosities, and the group Sp represents the magnitude of spin viscosity relative to the stress-spin coupling (analogous to the rotational viscosity in equilibrium systems). We assume that the walls of the cylinder are sufficiently rough that no-slip boundary conditions are enforced for the velocity,

$$\tilde{v}_\theta(\xi) = 1 \text{ and } \tilde{v}_\theta(1) = 0, \qquad [144]$$

and zero-spin boundary conditions are enforced for spin,

$$\tilde{M}_z(\xi) = \tilde{M}_z(1) = 0. \qquad [145]$$

The general solutions to the above coupled second order ODEs are

$$\tilde{v}_\theta = \frac{A\tilde{r}}{2(1 - \tilde{\lambda}_3)} + \frac{B\tilde{\lambda}_3}{\beta}I_1(\beta\tilde{r}) - \frac{C\tilde{\lambda}_3}{\beta}K_1(\beta\tilde{r}) + \frac{D}{\tilde{r}}, \qquad [146]$$

$$\tilde{M}_z = \frac{A}{1 - \tilde{\lambda}_3} + BI_0(\beta\tilde{r}) + CK_0(\beta\tilde{r}), \qquad [147]$$

$$\tilde{J}_z = \tilde{T}_{\theta r}^\mathrm{A} = \frac{1}{\tilde{r}}\frac{d}{d\tilde{r}}(\tilde{r}\tilde{v}) - \tilde{M} = -(1 - \tilde{\lambda}_3)\left[BI_0(\beta\tilde{r}) + CK_0(\beta\tilde{r})\right], \qquad [148]$$

where $I_n$ and $K_n$ are the $n$th order modified Bessel functions of the first and second kind, respectively; $A, B, C, D$ are integration constants and

$$\beta := \frac{1}{\tilde{\ell}} = \sqrt{\frac{1 - \tilde{\lambda}_3}{\mathrm{Sp}}}. \qquad [149]$$

Note $\tilde{\ell} := 1/\beta$ sets the scale of a boundary layer where the total stress is asymmetric. Using Eq. [25] in the main text, we may also provide a molecular interpretation of this boundary layer

$$\tilde{\ell} = \frac{d_\mathrm{c}}{4R_\mathrm{o}}\sqrt{(\alpha_2 + \alpha_3)(\lambda_2^{-1} + \lambda_3^{-1})[x_0(1 - d_\mathrm{s}^2/d_\mathrm{c}^2) + d_\mathrm{s}^2/d_\mathrm{c}^2]}, \qquad [150]$$

with solute weight fraction $x_0 := m_\mathrm{c} c_0/\rho$. In dimensional units, this boundary layer is $\ell = \tilde{\ell} R_\mathrm{o}$. In many instances, as is the case with chiral bacteria in solution (23), $d_\mathrm{s}^2/d_\mathrm{c}^2 \ll 1$. Thus, in this limit, the nondimensional boundary layer $\tilde{\ell}$ is

$$\tilde{\ell} = \frac{d_\mathrm{c}}{4R_\mathrm{o}}\sqrt{x_0(\alpha_2 + \alpha_3)(\lambda_2^{-1} + \lambda_3^{-1})}. \qquad [151]$$

Thus, the asymmetric boundary layer depends on the size and concentration of the solute, and the ratio of the spin viscosities to viscosities.

Using the boundary conditions from Eqs. [144] and [145], we analytically solve Eqs. [146] and [147], plotted in Fig. 3(b), sweeping over various values of $\tilde{\ell}$. The corresponding integration constants are

$$\frac{A}{1 - \tilde{\lambda}_3} = \frac{2\beta\xi}{P}\left(I_0(\beta\xi)K_0(\beta) - I_0(\beta)K_0(\beta\xi)\right), \qquad [152]$$

$$B = \frac{2\beta\xi}{P}(K_0(\beta\xi) - K_0(\beta)), \qquad [153]$$

$$C = \frac{2\beta\xi}{R}(I_0(\beta\xi) - I_0(\beta)), \qquad [154]$$

$$D = \frac{\xi}{R}\left[-\frac{2\tilde{\lambda}_3}{\beta} + K_0(\beta\xi)\left[2\tilde{\lambda}_3 I_1(\beta) - \beta I_0(\beta)\right] + I_0(\beta\xi)\left[2\tilde{\lambda}_3 K_1(\beta) + \beta K_0(\beta)\right]\right], \qquad [155]$$

where $P$ and $R$ are defined as

$$P := \frac{4\tilde{\lambda}_3}{\beta} - 2\tilde{\lambda}_3\left[\xi I_1(\beta\xi)K_0(\beta) + I_1(\beta)K_0(\beta\xi)\right]$$
$$+ I_0(\beta\xi)\left[\beta(-1 + \xi^2)K_0(\beta) - 2\tilde{\lambda}_3 K_1(\beta)\right] + I_0(\beta)\left[-\beta(-1 + \xi^2)K_0(\beta\xi) - 2\tilde{\lambda}_3\xi K_1(\beta\xi)\right], \qquad [156]$$

$$R := -\frac{4\tilde{\lambda}_3}{\beta} + 2\tilde{\lambda}_3\xi\left[I_1(\beta\xi)K_0(\beta) + I_0(\beta)K_1(\beta\xi)\right]$$
$$+ K_0(\beta\xi)\left[\beta(-1 + \xi^2)I_0(\beta) + 2\tilde{\lambda}_3 I_1(\beta)\right] + I_0(\beta\xi)\left[\beta(1 - \xi^2)K_0(\beta) + 2\tilde{\lambda}_3 K_1(\beta)\right]. \qquad [157]$$

We may also consider the total current passing through the cylinder,

$$Q = \int_{R_\mathrm{i}}^{R_\mathrm{o}} J_z(r) 2\pi r \, dr. \qquad [158]$$

Nondimensionalizing this current as

$$\tilde{Q} = \frac{Q}{\psi c_0 \xi \Omega 2\pi R_\mathrm{o}^2} = \int_\xi^1 \tilde{J}_z(\tilde{r})\tilde{r}\, d\tilde{r}. \qquad [159]$$

and substituting the analytical solution, we obtain

$$\tilde{Q} = -\frac{1-\tilde{\lambda}_3}{\beta}\left[B(I_1(\beta) - \xi I_1(\xi\beta)) - C(K_1(\beta) - \xi K_1(\xi\beta))\right]. \qquad [160]$$

We find the following analytical expression f in the limit of $1/\beta = \tilde{\ell} \to 0$:

$$\lim_{\tilde{\ell}\to 0} \tilde{Q} = -\frac{2\xi(1-\tilde{\lambda}_3)}{1-\xi}\tilde{\ell}. \qquad [161]$$

Dimensionally,

$$\lim_{\tilde{\ell}\to 0} Q = -\frac{4\pi\xi^2(1-\tilde{\lambda}_3)\tilde{\ell}}{1-\xi}\psi c_0 R_o^2 \Omega. \qquad [162]$$

Thus in this regime, the total current is linear with the nondimensional boundary layer $\tilde{\ell}$, the viscodiffusive coefficient $\psi$, and the imposed mechanical rotation $\Omega$.

**Chiral Engine.** Now we present the analytical solution for the setup shown in Fig. 3(c), the chiral engine. For this case, we assume the ansatzes, $c = c(z)$, $\mathbf{v} = v_\theta(r)\mathbf{e}_\theta$, and $\mathbf{M} = M_z(r)\mathbf{e}_z$. With the dimensionless variables

$$\tilde{r} := r/R_o, \qquad [163]$$

$$\tilde{z} := z/L, \qquad [164]$$

$$\tilde{c} := \frac{c - c_L}{\Delta c}, \qquad [165]$$

$$\tilde{v}_\theta := \frac{v_\theta}{\phi \Delta c R_o/[(\lambda_2 + \lambda_3)L]} \qquad [166]$$

$$\tilde{M}_z := \frac{M_z}{\phi \Delta c/(\gamma L)}, \qquad [167]$$

$$\tilde{J}_z := \frac{J_z}{D\Delta c/L}, \qquad [168]$$

$$\tilde{T}_{\theta r} := \frac{T_{\theta r}}{\phi \Delta c/L}. \qquad [169]$$

$$\tilde{T}_{\theta r}^A := \frac{T_{\theta r} - T_{r\theta}}{\phi \Delta c/L}, \qquad [170]$$

where $\Delta c := c_H - c_L$, Eqs. [7]-[9] in the main text reduce to

$$0 = \mathrm{Cr}\left(\frac{\tilde{\lambda}_3}{\tilde{r}}\frac{d}{d\tilde{r}}(\tilde{r}\tilde{v}_\theta) - \tilde{M}_z\right)\frac{d\tilde{c}}{d\tilde{z}} + \frac{d^2\tilde{c}}{d\tilde{z}^2}, \qquad [171]$$

$$0 = \frac{d}{d\tilde{r}}\left(\frac{1}{\tilde{r}}\frac{d}{d\tilde{r}}(\tilde{r}\tilde{v}_\theta)\right) - \frac{d\tilde{M}_z}{d\tilde{r}}, \qquad [172]$$

$$0 = \frac{\mathrm{Sp}}{\tilde{r}}\frac{d}{d\tilde{r}}\left(\tilde{r}\frac{d\tilde{M}_z}{d\tilde{r}}\right) + \frac{\tilde{\lambda}_3}{\tilde{r}}\frac{d}{d\tilde{r}}(\tilde{r}\tilde{v}_\theta) - \tilde{M}_z + \frac{d\tilde{c}}{d\tilde{z}}, \qquad [173]$$

where $\tilde{\lambda}_3$ and Sp are defined identically to the case of the chiral generator,

$$\tilde{\lambda}_3 := \frac{\lambda_3}{\lambda_2 + \lambda_3}, \qquad [174]$$

$$\mathrm{Sp} := \frac{\alpha_2 + \alpha_3}{2\gamma R_o^2}, \qquad [175]$$

and

$$\mathrm{Cr} := \frac{\psi\phi\Delta c}{D\lambda_3}. \qquad [176]$$

As $\gamma$ might be concentration dependent, we consider the case where the concentration difference $\Delta c$ is sufficiently small compared to the magnitude of the chemical bath $c_L$, i.e. $\tilde{c}_L := c_L/\Delta c \gg 1$. Moreover, we consider the case that $\mathrm{Cr}\tilde{c}_L \ll 1$. Then, Eq. [171] becomes

$$\frac{d^2\tilde{c}}{d\tilde{z}^2} = 0, \qquad [177]$$

rendering the system of differential equations linear. With these simplifications, the nondimensional expressions for the diffusive flux, shear stress, and antisymmetric stress become

$$\tilde{J}_z = \mathrm{Cr}(\tilde{c} + \tilde{c}_L)\left(\frac{\tilde{\lambda}_3}{\tilde{r}}\frac{d}{d\tilde{r}}(\tilde{r}\tilde{v}_\theta) - \tilde{M}_z\right) - \frac{d\tilde{c}}{d\tilde{z}} \approx -\frac{d\tilde{c}}{d\tilde{z}}, \qquad [178]$$

$$\tilde{T}_{\theta r} = (1-\tilde{\lambda}_3)\tilde{r}\frac{d}{d\tilde{r}}\left(\frac{\tilde{v}_\theta}{\tilde{r}}\right) + \frac{\tilde{\lambda}_3}{\tilde{r}}\frac{d}{d\tilde{r}}(\tilde{r}\tilde{v}_\theta) - \tilde{M}_z + \frac{d\tilde{c}}{d\tilde{z}}, \qquad [179]$$

$$\tilde{T}_{\theta r}^A = \frac{\tilde{\lambda}_3}{\tilde{r}}\frac{d}{d\tilde{r}}(\tilde{r}\tilde{v}_\theta) - \tilde{M}_z + \frac{d\tilde{c}}{d\tilde{z}}. \qquad [180]$$

The axial chemical baths impose Dirichlet boundary conditions for the concentration field $\tilde{c}$. While the outer cylinder is held in place, we assume that no net forces act on the inner cylinder, thus corresponding to homogeneous Dirichlet and Neumann boundary conditions, respectively. Lastly, we assume zero-spin boundary conditions are valid because of sufficient friction with the wall. In summary, these lead to the following boundary conditions:

$$\tilde{c}(0) = 0 \text{ and } \tilde{c}(1) = 1, \qquad [181]$$

$$\tilde{T}_{\theta r}(\xi) = 0 \text{ and } \tilde{v}_\theta(1) = 0, \qquad [182]$$

$$\tilde{M}_z(\xi) = \tilde{M}_z(1) = 0, \qquad [183]$$

where once again $\xi := R_\mathrm{i}/R_\mathrm{o}$.

The general solution to the above ODEs is

$$\tilde{c} = E\tilde{z} + F \qquad [184]$$

$$\tilde{v}_\theta = \frac{A/\tilde{\lambda}_3 + E}{2(1-\tilde{\lambda}_3)}\tilde{r} + \frac{B}{\beta}I_1(\beta\tilde{r}) - \frac{C}{\beta}K_1(\beta\tilde{r}) + \frac{D}{\tilde{\lambda}_3 \tilde{r}}, \qquad [185]$$

$$\tilde{M}_z = \frac{A+E}{1-\tilde{\lambda}_3} + BI_0(\beta\tilde{r}) + CK_0(\beta\tilde{r}), \qquad [186]$$

$$\tilde{J}_z = -E, \qquad [187]$$

$$\tilde{T}_{\theta r} = 2(1-\tilde{\lambda}_3)\left[-B\frac{I_1(\beta\tilde{r})}{\beta\tilde{r}} + C\frac{K_1(\beta\tilde{r})}{\beta\tilde{r}} - \frac{D}{\tilde{\lambda}_3 \tilde{r}^2}\right], \qquad [188]$$

$$\tilde{T}_{\theta r}^\mathrm{A} = -(1-\tilde{\lambda}_3)[BI_0(\beta\tilde{r}) + CK_0(\beta\tilde{r})], \qquad [189]$$

with $A, B, C, D, E, F$ being the integration constants. Here, $\beta$ and $\tilde{\ell}$ are defined identically to the chiral generator, i.e.,

$$\beta := \frac{1}{\tilde{\ell}} = \sqrt{\frac{1-\tilde{\lambda}_3}{\mathrm{Sp}}}. \qquad [190]$$

Once again, invoking the molecular picture of Eq. [25] in the main text reduces the nondimensional boundary layer $\tilde{\ell}$ to

$$\tilde{\ell} = \frac{d_\mathrm{c}}{4R_\mathrm{o}}\sqrt{(\alpha_2 + \alpha_3)(\lambda_2^{-1} + \lambda_3^{-1})[x_\mathrm{L}(1-d_\mathrm{s}^2/d_\mathrm{c}^2) + d_\mathrm{s}^2/d_\mathrm{c}^2]}, \qquad [191]$$

with solute weight fraction $x_\mathrm{L} := m_\mathrm{c} c_\mathrm{L}/\rho$. In the limit that $d_\mathrm{s} \ll d_\mathrm{c}$,

$$\tilde{\ell} = \frac{d_\mathrm{c}}{4R_\mathrm{o}}\sqrt{x_\mathrm{L}(\alpha_2 + \alpha_3)(\lambda_2^{-1} + \lambda_3^{-1})}. \qquad [192]$$

With the boundary conditions given in Eqs. [181]-[183], the integration constants $A, B, C, D, E, F$ are:

$$A = \frac{\tilde{\lambda}_3(\beta(-2\xi K_0(\beta)I_1(\beta\xi) + \beta I_2(\beta)K_0(\beta\xi) - 2\xi I_0(\beta)K_1(\beta\xi) - \beta K_2(\beta)I_0(\beta\xi)) + 4)}{\beta P}, \qquad [193]$$

$$B = \frac{\beta(K_0(\beta\xi) - K_0(\beta))}{P}, \qquad [194]$$

$$C = \frac{\beta(I_0(\beta\xi) - I_0(\beta))}{R}, \qquad [195]$$

$$D = \left[\frac{\beta(2\tilde{\lambda}_3 I_1(\beta) - \beta I_0(\beta))K_0(\beta\xi) + \beta(2\tilde{\lambda}_3 K_1(\beta) + \beta K_0(\beta))I_0(\beta\xi) - 2\tilde{\lambda}_3}{\tilde{\lambda}_3(\beta\xi K_0(\beta)I_1(\beta\xi) + \beta\xi I_0(\beta)K_1(\beta\xi) - 1)} + 2\right]^{-1}, \qquad [196]$$

$$E = 1, \qquad [197]$$

$$F = 0, \qquad [198]$$

where $P$ and $R$ are defined as

$$P := 2\tilde{\lambda}_3\xi K_0(\beta)I_1(\beta\xi) + 2\tilde{\lambda}_3\xi I_0(\beta)K_1(\beta\xi) + (2\tilde{\lambda}_3 I_1(\beta) - \beta I_0(\beta))K_0(\beta\xi) + (2\tilde{\lambda}_3 K_1(\beta) + \beta K_0(\beta))I_0(\beta\xi) - \frac{4\tilde{\lambda}_3}{\beta}, \qquad [199]$$

$$R := -2\tilde{\lambda}_3(\xi K_0(\beta)I_1(\beta\xi) + I_1(\beta)K_0(\beta\xi) + \xi I_0(\beta)K_1(\beta\xi)) - (2\tilde{\lambda}_3 K_1(\beta) + \beta K_0(\beta))I_0(\beta\xi) + \frac{4\tilde{\lambda}_3}{\beta} + \beta I_0(\beta)K_0(\beta\xi). \qquad [200]$$

The above solution is plotted in Fig. 3(d), for various values of $\tilde{\ell}$.

We may calculate the resulting rotation rate of the inner cylinder as

$$\Omega = v_\theta(R_\mathrm{i})/R_\mathrm{i}, \qquad [201]$$

or nondimensionally,

$$\tilde{\Omega} = \frac{\Omega}{\phi\Delta c/[\xi(\lambda_2 + \lambda_3)L]} = \tilde{v}_\theta(\xi). \qquad [202]$$

In the limit of $1/\beta = \tilde{\ell} \to 0$,

$$\lim_{\tilde{\ell}\to 0}\tilde{\Omega} = \xi(1+\xi)\tilde{\ell}, \qquad [203]$$

which in dimensional form corresponds to

$$\lim_{\tilde{\ell}\to 0}\Omega = \frac{(1+\xi)\tilde{\ell}}{(\lambda_2 + \lambda_3)L}\phi\Delta c. \qquad [204]$$

Thus, in the regime where $\tilde{\ell}$ is small, the inner rotation $\Omega$ is linearly dependent on the nondimensional boundary layer $\tilde{\ell}$, viscodiffusive transport coefficient $\phi$, and concentration difference $\Delta c$.

## D. Chiral Drift of Immotile Bacteria under Poiseuille Flow: Numerical Details

**Problem Formulation and Non-Dimensionalization.** Here, we provide numerical details and supplementary information pertaining to Fig. 4, mirroring the experimental setup in Ref. (23). Our goal is to numerically solve for the concentration, velocity, and spin fields, using Eqs. [1]-[6], and [25] from the main text. A solution of chiral immotile bacteria in solvent lies in a channel with dimensions $x \in [0, L], y \in [0, H]$, and $z \in [0, W]$. The bacteria is injected into the system with some initial profile at $c(0, y, z)$. The fluid is set into motion with a body force $\rho b \mathbf{e}_x$, creating a Poiseuille flow due to no-slip boundary conditions at the wall. As the fluid is viscodiffusive in nature, the applied body force will lead to fluxes of the chiral bacteria. We assume that the walls of the channel are impermeable to the solute, enforcing the boundary condition $\mathbf{J} \cdot \mathbf{n} = 0$, and that the walls are sufficiently rough enough to enforce the no slip $\mathbf{v} = \mathbf{0}$ and zero-spin $\mathbf{M} = \mathbf{0}$ boundary conditions.

In practice, we solve for the angular velocity field $\boldsymbol{\omega}$ (with $I\boldsymbol{\omega} = \rho \mathbf{M}$) rather than the spin angular momentum field $\mathbf{M}$. We make the ansatzes $c = c(x, y, z)$, $\mathbf{v} = v_x(x, y, z)\mathbf{e}_x$, $\boldsymbol{\omega} = \omega_y(x, y, z)\mathbf{e}_y + \omega_z(x, y, z)\mathbf{e}_z$, where we neglect $v_y, v_z$, and $\omega_x$. We begin by non-dimensionalizing the coordinates as

$$\tilde{x} := x/L, \quad \tilde{y} := y/H, \quad \tilde{z} := z/W, \qquad [205]$$

and the fields as

$$\tilde{c} := c/c_0, \qquad [206]$$

$$\tilde{v} := \frac{v_x}{\rho b H^2/(2\lambda_2)} = \frac{v_x}{U}, \qquad [207]$$

$$\tilde{\omega}_y := \frac{\omega_y}{U/(2W)}, \qquad [208]$$

$$\tilde{\omega}_z := \frac{\omega_z}{U/(2H)}, \qquad [209]$$

$$\tilde{I} := \frac{I}{m_c c_0 d_c^2/4} = \tilde{c}\left[1 - \left(\frac{d_s}{d_c}\right)^2\right] + \frac{\rho}{m_c c_0}\left(\frac{d_s}{d_c}\right)^2. \qquad [210]$$

We make use of the following values for the transport coefficients:

| Quantity | Value | Units | Source |
|---|---|---|---|
| $L$ | $1.075 \times 10^6$ | $\mu$m | (23) |
| $H$ | $9 \times 10^1$ | $\mu$m | (23) |
| $W$ | $10^3$ | $\mu$m | (23) |
| $U$ | $1.8 \times 10^4$ | $\mu$m/s | (23) |
| $\lambda_2/\rho$ | $10^6$ | $\mu$m$^2$/s | (23) |
| $D$ | $5 \times 10^{-2}$ | $\mu$m$^2$/s | (23) |
| $\lambda_3/\lambda_2$ | $2 \times 10^{-1}$ | - | (34) |
| $\alpha_1/\lambda_3$ | $2.22 \times 10^2$ | - | Assn. identical to $\alpha_2$ and $\alpha_3$ |
| $\alpha_2/\lambda_3$ | $2.22 \times 10^2$ | - | (34); Unit conversion using $d_s$ |
| $\alpha_3/\lambda_3$ | $2.22 \times 10^2$ | - | (34); Unit conversion using $d_s$ |
| $\psi$ | $5 \times 10^{-3}$ | $\mu$m | Fitted to (23) |
| $m_c c_0/\rho$ | $10^{-1}$ | - | Assn. |
| $d_c$ | $1.6 \times 10^1$ | $\mu$m | (23) |
| $d_s$ | $3 \times 10^{-4}$ | $\mu$m | Water molecule diameter |
| $k_B T$ | $4.11 \times 10^{-6}$ | fJ=g$\cdot\mu$m$^2$/s$^2$ | Value at $T = 298$K |
| $s$ | 1 | - | Assn. |
| $m_c$ | $5 \times 10^{-13}$ | g | (23); Calculate from water density |

The two free parameters here are the viscodiffusive coefficient $\psi$ and solute weight fraction $m_c c_0/\rho$, the former we fit to experimental data while assuming a value for the latter. Using the above nondimensional definitions, the governing PDEs [7]-[9] from the main text become

$$\text{Pe}\left(\frac{H}{L}\right)^2 \tilde{v}\frac{\partial \tilde{c}}{\partial \tilde{x}} = \tilde{\boldsymbol{\nabla}}_{yz} \cdot (\mathbf{g}^2 \tilde{\boldsymbol{\nabla}}_{yz}\tilde{c}) - \text{Vi}_J \frac{H}{W}\left[(\tilde{\boldsymbol{\nabla}}_{yz} \times \tilde{v}\mathbf{e}_x - \tilde{\boldsymbol{\omega}}) \cdot \tilde{\boldsymbol{\nabla}}_{yz}\tilde{c} - \tilde{c}\tilde{\boldsymbol{\nabla}}_{yz} \cdot \tilde{\boldsymbol{\omega}}\right], \qquad [211]$$

$$\text{Re}\left(\frac{H}{L}\right)^2 \tilde{v}\frac{\partial \tilde{v}}{\partial \tilde{x}} = 2 + \tilde{\boldsymbol{\nabla}}_{yz} \cdot (\mathbf{g}^2 \tilde{\boldsymbol{\nabla}}_{yz}\tilde{v}) - \frac{\lambda_3}{\lambda_2}(\mathbf{g}\tilde{\boldsymbol{\nabla}}_{yz}) \times \left[\mathbf{g}(\tilde{\boldsymbol{\nabla}}_{yz} \times \tilde{v}\mathbf{e}_x - \tilde{\boldsymbol{\omega}})\right], \qquad [212]$$

$$\text{Re}_\omega \tilde{v}\frac{\partial \tilde{I}\tilde{\boldsymbol{\omega}}}{\partial \tilde{x}} = \text{Sp}_1 \mathbf{g}^2 \tilde{\boldsymbol{\nabla}}_{yz}(\tilde{\boldsymbol{\nabla}}_{yz} \cdot (\tilde{I}\tilde{\boldsymbol{\omega}})) + \text{Sp}(\tilde{\boldsymbol{\nabla}}_{yz} \cdot \mathbf{g}^2 \tilde{\boldsymbol{\nabla}}_{yz})(\tilde{I}\tilde{\boldsymbol{\omega}}) + \tilde{\boldsymbol{\nabla}}_{yz} \times \tilde{v}\mathbf{e}_x - \tilde{\boldsymbol{\omega}} + \text{Vi}_T \frac{W}{H}\mathbf{g}^2 \tilde{\boldsymbol{\nabla}}_{yz}\tilde{c}, \qquad [213]$$

with the following dimensionless groups

$$\text{Vi}_J := \frac{\psi U}{D} = 1.9 \times 10^3, \qquad [214]$$

$$\text{Vi}_T := \frac{\phi c_0}{\lambda_3 U} = \frac{k_B T \psi c_0}{\lambda_3 U s} = 1.2 \times 10^{-6}, \qquad [215]$$

$$\text{Sp}_1 := \frac{1}{16}\frac{m_c c_0}{\rho}\frac{\alpha_1 + \alpha_2/3 - \alpha_3}{\lambda_3}\left(\frac{d_c}{H}\right)^2 = 1.46 \times 10^{-2}, \qquad [216]$$

$$\text{Sp} := \frac{1}{16}\frac{m_c c_0}{\rho}\frac{\alpha_2 + \alpha_3}{\lambda_3}\left(\frac{d_c}{H}\right)^2 = 8.78 \times 10^{-2}, \qquad [217]$$

$$\text{Pe} := \frac{UL}{D} = 3.87 \times 10^{11}, \qquad [218]$$

$$\text{Re} := \frac{UL}{\lambda_2/\rho} = 1.94 \times 10^4, \qquad [219]$$

$$\mathrm{Re}_\omega := \frac{1}{16}\frac{m_\mathrm{c} c_0}{\rho}\frac{UL}{\lambda_{3/\rho}}\left(\frac{d_c}{L}\right)^2 = 1.34\times 10^{-7}\,, \qquad [220]$$

$$\mathbf{g} := \mathbf{e}_y\otimes\mathbf{e}_y + \frac{H}{W}\mathbf{e}_z\otimes\mathbf{e}_z = \mathbf{e}_y\otimes\mathbf{e}_y + 0.09\mathbf{e}_z\otimes\mathbf{e}_z\,, \qquad [221]$$

$$\left(\frac{H}{L}\right)^2 = 7.01\times 10^{-9}\,, \qquad [222]$$

$$\frac{H}{W} = 9\times 10^{-2}\,. \qquad [223]$$

Due to the large lengthscale in $x$, derivatives in $x$ only appear in advective/inertial terms. Now, further neglecting $\mathrm{Re}\left(\frac{H}{L}\right)^2$, $\mathrm{Re}_\omega$, and $\mathrm{Vi}_T \frac{W}{H}$ as they are $\mathcal{O}(10^{-4})$ or less, we obtain

$$\mathrm{Pe}\left(\frac{H}{L}\right)^2 \tilde{v}\frac{\partial\tilde{c}}{\partial\tilde{x}} = \tilde{\boldsymbol{\nabla}}_{yz}\cdot(\mathbf{g}^2\tilde{\boldsymbol{\nabla}}_{yz}\tilde{c}) - \mathrm{Vi}_J \frac{H}{W}\left[(\tilde{\boldsymbol{\nabla}}_{yz}\times\tilde{v}\mathbf{e}_x - \tilde{\boldsymbol{\omega}})\cdot\tilde{\boldsymbol{\nabla}}_{yz}\tilde{c} - \tilde{c}\tilde{\boldsymbol{\nabla}}_{yz}\cdot\tilde{\boldsymbol{\omega}}\right]\,, \qquad [224]$$

$$0 = 2 + \tilde{\boldsymbol{\nabla}}_{yz}\cdot(\mathbf{g}^2\tilde{\boldsymbol{\nabla}}_{yz}\tilde{v}) - \frac{\lambda_3}{\lambda_2}(\mathbf{g}\tilde{\boldsymbol{\nabla}}_{yz})\times\left[\mathbf{g}(\tilde{\boldsymbol{\nabla}}_{yz}\times\tilde{v}\mathbf{e}_x - \tilde{\boldsymbol{\omega}})\right]\,, \qquad [225]$$

$$\mathbf{0} = \mathrm{Sp}_1 \mathbf{g}^2 \tilde{\boldsymbol{\nabla}}_{yz}(\tilde{\boldsymbol{\nabla}}_{yz}\cdot(\tilde{I}\tilde{\boldsymbol{\omega}})) + \mathrm{Sp}(\tilde{\boldsymbol{\nabla}}_{yz}\cdot\mathbf{g}^2\tilde{\boldsymbol{\nabla}}_{yz})(\tilde{I}\tilde{\boldsymbol{\omega}}) + \tilde{\boldsymbol{\nabla}}_{yz}\times\tilde{v}\mathbf{e}_x - \tilde{\boldsymbol{\omega}}\,, \qquad [226]$$

where we see that the system is in the Stokes flow regime—the velocity and spin fields have negligible inertial transport. These PDEs are solved with the boundary conditions $\tilde{\mathbf{J}}\cdot\mathbf{n} = 0$, $\tilde{\boldsymbol{\omega}} = \mathbf{0}$, and $\tilde{v} = 0$. An concentration of $\tilde{c}(0,\tilde{y},\tilde{z}) = \tilde{c}_{\mathrm{IC}}(\tilde{z})$ is used as the initial condition, scaled to 1 by $c_0$. Specifically, we use

$$\tilde{c}_{\mathrm{IC}}(\tilde{z}) = \epsilon + \Theta(\tilde{z}-0.45) - \Theta(\tilde{z}-0.55)\,, \qquad [227]$$

with Heaviside function $\Theta(\tilde{z})$ and where the parameter $\epsilon = 10^{-2}$ is introduced to confer stability when numerically solving the nonlinear problem.

**Weak Form.** We now proceed to solve Eqs. [224]-[226] using the finite element method (FEM). To that end, given the nature of the equations, we pose a boundary value problem (BVP) in the $\tilde{y}\tilde{z}$ plane and an initial value problem (IVP) in the $\tilde{x}$ direction. To obtain the weak form, we first multiply Eqs. [224]-[226] by the test functions $\delta c$, $\delta v$, $\delta\omega_y$ and $\delta\omega_z$, integrate over the domain $\Omega = [0,1]\times[0,1]$ on the $\tilde{y}\tilde{z}$ plane, using integration by parts and obtain

$$\mathrm{Pe}\left(\frac{H}{L}\right)^2 \int_\Omega \tilde{v}\frac{\partial\tilde{c}}{\partial\tilde{x}}\delta c\,d\Omega - \mathrm{Vi}_J \int_\Omega \tilde{\boldsymbol{\nabla}}_{yz}\delta c\cdot(\mathbf{g}\tilde{\mathbf{J}})\,d\Omega = 0\,, \qquad [228]$$

$$\int_\Omega \tilde{\boldsymbol{\nabla}}_{yz}\delta v\cdot\mathbf{g}\left(\tilde{\mathbf{t}}^\mathrm{S} + \frac{\lambda_3}{\lambda_2}\tilde{\mathbf{t}}^\mathrm{A}\right)d\Omega - 2\int_\Omega \delta v\,d\Omega = 0\,, \qquad [229]$$

$$\mathrm{Sp}_1 \int_\Omega \frac{\partial\delta\omega_y}{\partial\tilde{y}}\tilde{\boldsymbol{\nabla}}_{yz}\cdot(\tilde{I}\tilde{\boldsymbol{\omega}})\,d\Omega + \mathrm{Sp}\int_\Omega \tilde{\boldsymbol{\nabla}}_{yz}\delta\omega_y\cdot\mathbf{g}^2\tilde{\boldsymbol{\nabla}}_{yz}(\tilde{I}\tilde{\omega}_y)\,d\Omega - \frac{W}{H}\int_\Omega \delta\omega_y \tilde{\mathbf{t}}^\mathrm{A}\cdot\mathbf{e}_z\,d\Omega = 0\,, \qquad [230]$$

$$\mathrm{Sp}_1\left(\frac{H}{W}\right)^2\int_\Omega \frac{\partial\delta\omega_z}{\partial\tilde{z}}\tilde{\boldsymbol{\nabla}}_{yz}\cdot(\tilde{I}\tilde{\boldsymbol{\omega}})\,d\Omega + \mathrm{Sp}\int_\Omega\left[\tilde{\boldsymbol{\nabla}}_{yz}\delta\omega_z\cdot\mathbf{g}^2\tilde{\boldsymbol{\nabla}}_{yz}(\tilde{I}\tilde{\omega}_z) + \left(\frac{H}{W}\right)^2 \frac{\xi_\omega h}{2}\frac{\partial\delta\omega_z}{\partial\tilde{z}}\frac{\partial\tilde{\omega}_z}{\partial\tilde{z}}\right]d\Omega + \int_\Omega \delta\omega_z \tilde{\mathbf{t}}^\mathrm{A}\cdot\mathbf{e}_y\,d\Omega = 0\,, \qquad [231]$$

where

$$\tilde{\mathbf{J}} := -\frac{1}{\mathrm{Vi}_J}\mathbf{g}\tilde{\boldsymbol{\nabla}}_{yz}\tilde{c} + \left(\frac{H}{W}\right)\tilde{c}\left(\frac{\partial\tilde{v}}{\partial\tilde{z}} - \tilde{\omega}_y\right)\mathbf{e}_y - \tilde{c}\left(\frac{\partial\tilde{v}}{\partial y} + \tilde{\omega}_z\right)\mathbf{e}_z - \frac{\xi_c h}{2}\frac{\partial\tilde{c}}{\partial\tilde{z}}\mathbf{e}_z\,, \qquad [232]$$

$$\tilde{\mathbf{t}}^\mathrm{S} := \mathbf{g}\tilde{\boldsymbol{\nabla}}\tilde{v} = \frac{\partial\tilde{v}}{\partial\tilde{y}}\mathbf{e}_y + \frac{H}{W}\frac{\partial\tilde{v}}{\partial\tilde{z}}\mathbf{e}_z\,, \qquad [233]$$

$$\tilde{\mathbf{t}}^\mathrm{A} := \left[\frac{\partial\tilde{v}}{\partial\tilde{y}} + \tilde{\omega}_z - \mathrm{Vi}_T\frac{H}{W}\frac{\partial\tilde{c}}{\partial\tilde{z}}\right]\mathbf{e}_y + \frac{H}{W}\left[\frac{\partial\tilde{v}}{\partial\tilde{z}} - \tilde{\omega}_y + \mathrm{Vi}_T\frac{W}{H}\frac{\partial\tilde{c}}{\partial\tilde{y}}\right]\mathbf{e}_z\,, \qquad [234]$$

$h$ is the maximum mesh size and the parameters $\xi_c$ and $\xi_\omega$ stabilize the convection-dominated problem by adding numerical diffusion. We set $\xi_c = \xi_\omega = 2$. Moreover, we have used the boundary conditions $\mathbf{J}\cdot\mathbf{n} = 0$ and $\delta v = \delta\omega_y = \delta\omega_z = 0$ to eliminate boundary terms.

For the numerical scheme, we use the backward Euler method to integrate in $x$, leading to

$$\mathrm{Pe}\left(\frac{H}{L}\right)^2 \int_\Omega \tilde{v}_n \frac{\tilde{c}_n - \tilde{c}_{n-1}}{\Delta\tilde{x}}\delta c\,d\Omega - \mathrm{Vi}_J\int_\Omega \tilde{\boldsymbol{\nabla}}_{yz}\delta c\cdot(\mathbf{g}\tilde{\mathbf{J}}_n)\,d\Omega = 0\,, \qquad [235]$$

$$\int_\Omega \tilde{\boldsymbol{\nabla}}_{yz}\delta v\cdot\mathbf{g}\left(\tilde{\mathbf{t}}_n^\mathrm{S} + \frac{\lambda_3}{\lambda_2}\tilde{\mathbf{t}}_n^\mathrm{A}\right)d\Omega - 2\int_\Omega \delta v\,d\Omega = 0\,, \qquad [236]$$

$$\mathrm{Sp}_1 \int_\Omega \frac{\partial\delta\omega_y}{\partial\tilde{y}}\tilde{\boldsymbol{\nabla}}_{yz}\cdot(\tilde{I}_n\tilde{\boldsymbol{\omega}}_n)\,d\Omega + \mathrm{Sp}\int_\Omega \tilde{\boldsymbol{\nabla}}_{yz}\delta\omega_y\cdot\mathbf{g}^2\tilde{\boldsymbol{\nabla}}_{yz}(\tilde{I}_n\tilde{\omega}_{yn})\,d\Omega - \frac{W}{H}\int_\Omega \delta\omega_y \tilde{\mathbf{t}}_n^\mathrm{A}\cdot\mathbf{e}_z\,d\Omega = 0\,, \qquad [237]$$

$$\mathrm{Sp}_1\left(\frac{H}{W}\right)^2\int_\Omega \frac{\partial\delta\omega_z}{\partial\tilde{z}}\tilde{\boldsymbol{\nabla}}_{yz}\cdot(\tilde{I}_n\tilde{\boldsymbol{\omega}}_n)\,d\Omega + \mathrm{Sp}\int_\Omega\left[\tilde{\boldsymbol{\nabla}}_{yz}\delta\omega_z\cdot\mathbf{g}^2\tilde{\boldsymbol{\nabla}}_{yz}(\tilde{I}_n\tilde{\omega}_{zn}) + \left(\frac{H}{W}\right)^2 \frac{\xi_\omega h}{2}\frac{\partial\delta\omega_z}{\partial\tilde{z}}\frac{\partial\tilde{\omega}_{zn}}{\partial\tilde{z}}\right]d\Omega + \int_\Omega \delta\omega_z \tilde{\mathbf{t}}_n^\mathrm{A}\cdot\mathbf{e}_y\,d\Omega = 0\,, \qquad [238]$$

$$\tilde{\mathbf{J}}_n := -\frac{1}{\mathrm{Vi}_J}\mathbf{g}\tilde{\boldsymbol{\nabla}}_{yz}\tilde{c}_n + \left(\frac{H}{W}\right)\tilde{c}_n\left(\frac{\partial\tilde{v}_n}{\partial\tilde{z}} - \tilde{\omega}_{yn}\right)\mathbf{e}_y - \tilde{c}_n\left(\frac{\partial\tilde{v}_n}{\partial y} + \tilde{\omega}_{zn}\right)\mathbf{e}_z - \frac{\xi_c h}{2}\frac{\partial\tilde{c}_n}{\partial\tilde{z}}\mathbf{e}_z\,, \qquad [239]$$

$$\tilde{\mathbf{t}}_n^\mathrm{S} := \mathbf{g}\tilde{\boldsymbol{\nabla}}\tilde{v}_n = \frac{\partial\tilde{v}_n}{\partial\tilde{y}}\mathbf{e}_y + \frac{H}{W}\frac{\partial\tilde{v}_n}{\partial\tilde{z}}\mathbf{e}_z\,, \qquad [240]$$

$$\tilde{\mathbf{t}}_n^\mathrm{A} := \left[\frac{\partial\tilde{v}_n}{\partial\tilde{y}} + \tilde{\omega}_{zn} - \mathrm{Vi}_T\frac{H}{W}\frac{\partial\tilde{c}_n}{\partial\tilde{z}}\right]\mathbf{e}_y + \frac{H}{W}\left[\frac{\partial\tilde{v}_n}{\partial\tilde{z}} - \tilde{\omega}_{yn} + \mathrm{Vi}_T\frac{W}{H}\frac{\partial\tilde{c}_n}{\partial\tilde{y}}\right]\mathbf{e}_z\,, \qquad [241]$$

where $f_n$ denotes the numerical estimate for $f(x_n)$. We use the finite element package **FEniCSx** to solve, choosing continuous first-order Lagrange basis functions on an unstructured triangular mesh generated with **gmsh**. Starting with the initial condition $\tilde{c}(0,\tilde{y},\tilde{z}) = \tilde{c}_{\mathrm{IC}}(\tilde{z})$, we use a temporal step of $\Delta\tilde{x} = 1500$ and a spatial mesh size of $h = 0.0025$ to report the results presented in Fig. 4 of the main text.